\documentclass[5p,times]{cas-dc}

\usepackage[square,sort,comma,numbers]{natbib}

\usepackage[cmex10]{amsmath}

\usepackage{array}
\usepackage{mdwmath}
\usepackage{mdwtab}
\usepackage{eqparbox}
\usepackage{url}
\usepackage{supertabular}                                             
\usepackage{amssymb}
\usepackage{caption}
\usepackage{subcaption}
\usepackage{float}
\usepackage{booktabs}
\usepackage{multirow}
\usepackage{graphicx}
\usepackage{color}
\usepackage{tabularray}
\usepackage{hhline}
\usepackage{wasysym}
\usepackage[ruled, lined, longend, linesnumbered]{algorithm2e}

\setcitestyle{square}
\def\tsc#1{\csdef{#1}{\textsc{\lowercase{#1}}\xspace}}
\tsc{WGM}
\tsc{QE}
\tsc{EP}
\tsc{PMS}
\tsc{BEC}
\tsc{DE}

\begin{document}
\let\WriteBookmarks\relax
\def\floatpagepagefraction{1}
\def\textpagefraction{.001}

\shorttitle{Deep Reinforcement Learning-based Scheduling in Edge and Fog Computing Environments}

\shortauthors{Wang et~al.}

\title [mode = title]{Deep Reinforcement Learning-based Scheduling for Optimizing System Load and Response Time in Edge and Fog Computing Environments}                      

\author[1]{Zhiyu Wang}
\ead{zhiywang1@student.unimelb.edu.au}

\affiliation[1]{organization={Cloud Computing and Distributed Systems (CLOUDS) Laboratory, The University of Melbourne},
    city={Melbourne},
    country={Australia}}

\author[2]{Mohammad Goudarzi}

\affiliation[2]{organization={School of Computer Science and Engineering, The University of New South Wales (UNSW)},
    city={Sydney},
    country={Australia}}
\ead{m.goudarzi@unsw.edu.au}

\author[3]{Mingming Gong}
\ead{mingming.gong@unimelb.edu.au}

\author[1]{Rajkumar Buyya}
\ead{rbuyya@unimelb.edu.au}

\affiliation[3]{organization={School of Mathematics and Statistics, The University of Melbourne},
    city={Melbourne},
    country={Australia}}

\begin{abstract}
Edge/fog computing, as a distributed computing paradigm, satisfies the low-latency requirements of ever-increasing number of IoT applications and has become the mainstream computing paradigm behind IoT applications. However, because large number of IoT applications require execution on the edge/fog resources, the servers may be overloaded. Hence, it may disrupt the edge/fog servers and also negatively affect IoT applications' response time. Moreover, many IoT applications are composed of dependent components incurring extra constraints for their execution. Besides, edge/fog computing environments and IoT applications are inherently dynamic and stochastic. Thus, efficient and adaptive scheduling of IoT applications in heterogeneous edge/fog computing environments is of paramount importance. However, limited computational resources on edge/fog servers imposes an extra burden for applying optimal but computationally demanding techniques. To overcome these challenges, we propose a Deep Reinforcement Learning-based IoT application Scheduling algorithm, called DRLIS to adaptively and efficiently optimize the response time of heterogeneous IoT applications and balance the load of the edge/fog servers. We implemented DRLIS as a practical scheduler in the FogBus2 function-as-a-service framework for creating an edge-fog-cloud integrated serverless computing environment. Results obtained from extensive experiments show that DRLIS significantly reduces the execution cost of IoT applications by up to 55\%, 37\%, and 50\% in terms of load balancing, response time, and weighted cost, respectively, compared with metaheuristic algorithms and other reinforcement learning techniques. 
\end{abstract}


\maketitle

\section{Introduction}
The past few years have witnessed the rapid rise of the Internet of Things (IoT) industry, enabling the connection of people to things and things to things, and facilitating the digitization of the physical world \cite{chalapathi2021industrial}. Meanwhile, with the explosive growth of IoT devices and various applications, the expectation for stability and low latency is higher than ever \cite{azizi2022deadline}. As the main enabler of IoT, cloud computing stores and processes data and information generated by IoT devices. Leveraging powerful computing capabilities and advanced storage technologies, cloud computing ensures the security and reliability of stored information. However, servers in the cloud computing paradigm are usually located at a long physical distance from IoT devices, and the high latency caused by long distances cannot efficiently satisfy real-time IoT applications. Prompted by these issues, edge and fog computing computing have emerged as popular computing paradigms in the IoT context. Although some researchers use the terms edge computing and fog computing interchangeably, we clearly define them in this paper. We consider the case that use “only” edge resources for real-time IoT applications as edge computing, and the case that use edge and whenever necessary also utilizes cloud resources (along with edge resources in a seamless manner) as fog computing. Edge computing as a decentralized computing architecture brings processing, storage, and intelligent control to the vicinity of IoT devices \cite{ferrer2019towards}. This flexible architecture extends cloud computing services to the edge of the network. In contrast, the fog computing paradigm inherits the advantages of both cloud and edge computing \cite{goudarzi2020application}, which not only provides powerful computational capabilities but also reduces the need to transfer data to the cloud for processing, analysis, and storage, thus reducing the inter-network distance. In the real world, edge and fog computing provide strong support for innovation and development in various fields. For example, in the field of smart healthcare, deploying edge computing nodes on wearable devices and medical devices can monitor patients' physiological parameters in real time and transmit the data to the cloud for analysis and diagnosis, realizing telemedicine and personalized medicine \cite{7070665}; in the field of autonomous driving, deploying edge computing nodes on self-driving vehicles can perform real-time sensing and decision processing, enabling shorter response time and improving driving safety \cite{9288755}.

However, the massive growth in the number of IoT applications and servers in fog computing environments also creates new challenges. Firstly, the execution time is expected to be minimized \cite{goudarzi2021distributed}, which means that the applications should be processed by the best (i.e., the most powerful and physically closest) server. Besides, the load should be ideally balanced and distributed to run on multiple operating units. For example, by distributing requests across multiple servers in a seamless manner (as in serverless computing environments), load balancing can avoid overloading individual servers and ensure that each server handles a moderate load. This improves response times, overall system performance, and throughput, and also helps servers run more consistently. Therefore, improving the load balancing level of servers (i.e., lowering the variance of server resource utilization) while reducing the response time becomes an important but challenging problem for scheduling IoT applications on servers in edge/fog computing environments. Since this is an NP-hard problem, metaheuristic and rule-based solutions can be considered \cite{brogi2017qos}, \cite{goudarzi2022scheduling}. However, these approaches often rely on omniscient knowledge of global information and require the solution proponent to have control over the changes. In the fog computing environment, there is often no regularity in server performance, utilization, and downtime. The number of IoT applications and the corresponding resource requirements are even more nearly random. Besides, in reality, Directed Acyclic Graphs (DAGs) are often used to model IoT applications \cite{ma2019iot}, where nodes represent tasks and edges represent data communication between dependent tasks. The dependency among tasks introduces higher complexity in scheduling applications. Therefore, metaheuristic and rule-based solutions cannot efficiently cope with the IoT application scheduling problem in fog computing environments.

Deep Reinforcement Learning (DRL) is the product of combining deep learning with reinforcement learning, integrating the powerful understanding of deep learning on perceptual problems with the decision-making capabilities of reinforcement learning. In deep reinforcement learning, the agent continuously interacts with the environment, recording a large number of empirical trajectories (i.e., sequences of states, actions, and rewards), which are used in the training phase to learn optimal policies. In contrast to metaheuristic algorithms, agents in deep reinforcement learning are able to autonomously sense and respond to changes in the environment, which allows deep reinforcement learning to solve complex problems in realistic scenarios. However, due to the limited computational resources of devices in fog computing environments \cite{wang2022container}, the computational requirements of complex Deep Neural Networks (DNNs) are often not supported \cite{li2018edge}. Therefore, how to balance implementation simplicity, sample complexity, and solution performance becomes a key research problem in applying deep reinforcement learning to fog computing environments to cope with complex situations.

To address the above challenges, we propose a Deep Reinforcement Learning-based IoT application Scheduling algorithm (DRLIS), which employs Proximal Policy Optimization (PPO) \cite{schulman2017proximal} technique for solving the IoT applications scheduling problem in fog computing environments. DRLIS can effectively optimize the load balancing cost of the servers, the response time cost of the IoT applications, and their weighted cost. Besides, by using clipped surrogate objective to limit the magnitude of policy updates in each iteration and being able to perform multiple iterations of updates in the sampled data, the convergence speed of the algorithm is improved. Moreover, considering the limited computational resources and the optimization objective under study, we design efficient reward functions.
The main contributions of this paper are:
\begin{itemize}
\item We propose a weighted cost model regarding DAG-based IoT applications' scheduling in fog computing environments to improve the load balancing level of the servers while minimizing the response time of the application. In addition, we adapt this weighted cost model to make it applicable to DRL algorithms.
\item We propose a DRL-based algorithm (DRLIS) to solve the defined weighted cost optimization problem in dynamic and stochastic fog computing environments. When the computing environment changes (e.g., requests from different IoT applications, server computing resources, the number of servers), it can adaptively update the scheduling policy with a fast convergence speed.
\item Based on DRLIS, we implement a practical scheduler in the FogBus2 function-as-a-service framework\footnote{Please refer to \cite{deng2021fogbus2, goudarzi2021resource} for detailed description of the FogBus2 framework} \cite{deng2021fogbus2} for handling scheduling requests of IoT applications in heterogeneous fog and edge computing environments. We also extend the functionality of the FogBus2 framework to make different DRL techniques applicable to it.
\item We conduct practical experiments and use real IoT applications with heterogeneous tasks and resource demands to evaluate the performance of DRLIS in real system setup. By comparing with common metaheuristics (Non-dominated Sorting Genetic Algorithm 2 (NSGA2) \cite{deb2000fast}, Non-dominated Sorting Genetic Algorithm 3 (NSGA3) \cite{deb2013evolutionary}) and other reinforcement learning algorithms (Q-Learning \cite{watkins1992q}), we demonstrate the superiority of DRLIS in terms of convergence speed, optimization cost, and scheduling time.
\end{itemize}

The rest of the paper is organized as follows. Section \ref{related_work} discusses related work and Section \ref{system_model} presents the system model and problem formulation. The Deep Reinforcement Learning model for IoT applications in edge and fog computing environments is presented in Section \ref{drl_model}. DRLIS is discussed in Section \ref{ppo_algorithm}. Section \ref{performance_evaluation} evaluates the performance of DRLIS and compares it with other counterparts. Finally, Section \ref{conclusion} concludes the paper and states future work.

\section{Related Work}
\label{related_work}
In this section, we review the literature on scheduling IoT applications in edge and fog computing environments. The related works are divided into metaheuristic and reinforcement learning categories.

\subsection{Metaheuristic}
In the dependent category, Liu et al. \cite{7541539} adopted a Markov Decision Process (MDP) approach to achieving shorter average task execution latency in edge computing environments. They proposed an efficient one-dimensional search algorithm to find the optimal task scheduling policy. However, this work cannot adapt to changes in the computing environment and is difficult to extend to solve complex weighted cost optimization problems in heterogeneous fog computing environments. Wu et al. \cite{8587216} modeled the task scheduling problem in edge and fog computing environments as a DAG and used an estimation of distribution algorithm (EDA) and a partitioning operator to partition the graph in order to queue tasks and assign appropriate servers. However, they did not practically implement and test their work. Sun et al. \cite{sun2018multi} improved the NSGA2 algorithm and designed a resource scheduling scheme among fog nodes in the same fog cluster, taking into account the diversity of different devices. This work aims to reduce the service latency and improve the stability of task execution. Although capable of handling weighted cost optimization problems, this work only considers scheduling problems in the same computing environment. Hoseiny et al. \cite{9484436} proposed a Genetic Algorithm (GA)-based technique for minimizing the total computation time and energy consumption of task scheduling in a heterogeneous fog cloud computing environment. By introducing features for tasks, the technique can find a more suitable computing environment for each task. However, it does not consider the dependencies of different tasks in the application, and due to the use of metaheuristic algorithms, scheduling rules need to be manually set, which cannot adapt to changing computing environments. Ali et al. \cite{9233987} proposed an NSGA2-based technique for minimizing the total computation time and system cost of task scheduling in heterogeneous fog cloud computing environments. Their work formulates the task scheduling problem as an optimization problem in order to dynamically allocate appropriate resources for predefined tasks. Similarly, due to the limitations of metaheuristic algorithms, this work requires the assumption that the technique has some knowledge of the submitted tasks to develop the scheduling policy and thus cannot cope with dynamic and complex scenarios.

\subsection{Reinforcement Learning}
In the dependent category, Shahidani et al. \cite{ramezani2023task} proposed a Q-learning-based algorithm to reduce task execution latency and balance the load in a fog cloud computing environment. However, this work does not consider the inter-task dependencies and the heterogeneity of fog and cloud computing environments. Baek et al. \cite{baek2019managing} adapted the Q-learning algorithm and proposed an approach that aims at improving load balancing in fog computing environments. This work considers the heterogeneity of nodes in fog computing environments but still assumes that the tasks within the application are independent of each other. Jie et al. \cite{jie2021dqn} proposed a Deep Q-Network (DQN)-based approach to minimize the total latency of task processing in edge computing environments. This work formulates task scheduling as a Markov Decision Process while considering the heterogeneity of IoT applications. However, this work only considers the scheduling problem in edge computing environments and investigates only one optimization objective. Xiong et al. \cite{9060882} adapted the DQN algorithm and proposed a resource allocation strategy for IoT edge computing systems. This work aims at minimizing the average job completion time but does not take into account more complex functions with multiple optimization objectives. Wang et al. \cite{8657791} focus on edge computing environments and propose a deep reinforcement learning-based resource allocation (DRLRA) scheme based on DQN. This work targets to reduce the average service time and balance the resource usage within the edge computing environment. However, the work does not consider the resources in fog computing environment, and the technique is not practically implemented and tested. Huang et al. \cite{huang2019deep} adopted a DQN-based approach to address the resource allocation problem in the edge computing environment. This work investigated minimizing the weighted cost, including the total energy consumption and the latency to complete the task. However, it does not consider the heterogeneity of servers in fog computing environments and assumes that the tasks are independent. Chen et al. \cite{8493155} proposed an approach based on double DQN to balance task execution time and energy consumption in edge computing environments. Similarly, this work is only applicable to the edge environment and does not consider the dependencies between tasks. Zheng et al. \cite{9798187} proposed a Soft Actor-Critic (SAC)-based algorithm to minimize the task completion time in an edge computing environment. This work focuses on the latency problem and the experiments are simulation-based. Zhao et al. \cite{10136736} proposed a Twin Delayed DDPG (TD3)-based DRL algorithm. The goal of this work is to minimize the latency and energy consumption, but inter-task dependencies are not considered and the results are also simulation-based. Liao et al. \cite{liao2023online} used Deep Deterministic Policy Gradient (DDPG) and Double Deep Q-Network (DQN) algorithms to model computation in an edge environment. This work aims to reduce energy consumption and latency but does not consider the fog environment and the heterogeneity of devices. Sethi et al. \cite{sethi2023feddove} proposed a DQN-based algorithm to optimize energy consumption and load balancing of fog servers. Similarly, this work is simulation-based and does not consider the dependencies between tasks. 

Table \ref{tab:comparison} presents the comparison of the related work with our proposed algorithm, in terms of application properties, architecture properties, algorithm properties, and evaluation. In the application properties section, the number of tasks included in the IoT application, and the dependencies between tasks are studied. In the architectural properties section, three aspects are studied including the IoT device layer, the edge/fog layer, and the multi-cloud layer. For the IoT device layer, the application type and request type are identified. The real application section indicates that the work either deploys actual IoT applications, adopts simulated applications, or uses random data. The heterogeneous request type represents work considering that different IoT devices have different numbers of requests and different requirements. For the edge/fog layer, the computing environment and the heterogeneity of deployed servers are investigated. Besides, the multi-cloud layer studies whether the work considers the scenario of different cloud service providers with heterogeneity. In the algorithm properties section, we investigate the main technique on which each work is based and the corresponding optimization objectives. The evaluation section identifies whether the work is based on simulation or practical experiments. Recent works that we reviewed (e.g., \cite{9798187}, \cite{10136736}, \cite{liao2023online}, \cite{sethi2023feddove}, \cite{li2022deep}, \cite{9945614}, \cite{xue2022deep}) have often used reinforcement learning approaches to deal with workload scheduling problems. This is because reinforcement learning can learn by interacting with the environment and continuously optimizing the policy through feedback signals (e.g., reward or penalty). This learning ability gives reinforcement learning an advantage when facing complex, dynamic environments  \cite{pallewatta2023placement}, whereas metaheuristic techniques require manual adaptation and guidance.

\renewcommand{\arraystretch}{2}
\begin{table*}
\centering
\caption{A qualitative comparison of related works with ours}
\label{tab:comparison}
\resizebox{\textwidth}{!}{%
\begin{tabular}{|c|c|c|c|c|c|c|c|c|c|c|c|c|c|} 
\hline
\multirow{3}{*}{Works} & \multicolumn{2}{c|}{Application Properties}                 & \multicolumn{5}{c|}{Architectural Properties}                                                                       & \multicolumn{5}{c|}{Algorithm Properties}                                                                                                                     & \multirow{3}{*}{Evaluation}  \\ 
\cline{2-13}
                       & \multirow{2}{*}{Task Number} & \multirow{2}{*}{Dependency}  & \multicolumn{2}{c|}{IoT Device Layer} & \multicolumn{2}{c|}{Edge/Fog Layer}    & \multirow{2}{*}{Multi-Cloud Layer} & \multicolumn{2}{c|}{\multirow{2}{*}{Main Technique}}                                                          & \multicolumn{3}{c|}{Optimization Objectives}  &                              \\ 
\cline{4-7}\cline{11-13}
                       &                              &                              & Real Applications & Request Type      & Computing Environments & Heterogeneity &                                    & \multicolumn{2}{c|}{}                                                                                         & Time       & Load Balancing & Weighted &                              \\ 
\hline
 \cite{7541539}                      & Single                       & \multirow{4}{*}{Independent} & \Circle          & Homogeneous       & Edge                   & Homogeneous   & $\times$                           & \multirow{6}{*}{\begin{tabular}[c]{@{}c@{}}Metaheuristic \\ Algorithms\end{tabular}}                 & MDP        & \checkmark & $\times$       & $\times$        & Simulation                   \\ 
\cline{1-2}\cline{4-8}\cline{10-14}
 \cite{sun2018multi}                      & Multiple                     &                              & \LEFTcircle          & Homogeneous       & Edge and Fog           & Heterogeneous & $\times$                           &                                                                                                  & NSGA2      & \checkmark & $\times$       & \checkmark      & Simulation                   \\ 
\cline{1-2}\cline{4-8}\cline{10-14}
 \cite{9484436}                      & Single                       &                              & \Circle          & Homogeneous       & Edge and Fog           & Heterogeneous & $\times$                           &                                                                                                  & GA         & \checkmark & $\times$       & $\times$        & Simulation                   \\ 
\cline{1-2}\cline{4-8}\cline{10-14}
 \cite{9233987}                      & Single                       &                              & \Circle          & Homogeneous       & Edge and Fog           & Heterogeneous & $\times$                           &                                                                                                  & NSGA2      & \checkmark & $\times$       & \checkmark      & Simulation                   \\ 
\cline{1-8}\cline{10-14}
  \cite{8587216}                     & Multiple                       &    Dependent                          & \LEFTcircle          & Homogeneous       & Edge and Fog           & Heterogeneous   & $\times$                           &                                                                                                  & EDA        & \checkmark & $\times$       & \checkmark        & Simulation                   \\ 
\hline
  \cite{baek2019managing}                     & Single                       & \multirow{13}{*}{Independent} & \Circle          & Homogeneous       & Edge and Fog           & Heterogeneous & $\times$                           & \multirow{14}{*}{\begin{tabular}[c]{@{}c@{}}Reinforcement \\ Learning \\ Techniques\end{tabular}} & Q-Learning & $\times$   & \checkmark     & $\times$        & Simulation                   \\ 
\cline{1-2}\cline{4-8}\cline{10-14}
    \cite{ramezani2023task}                   & Single                       &                              & \Circle          & Homogeneous     & Edge and Fog                   & Homogeneous   & $\times$                           &                                                                                                  & Q-Learning & \checkmark & \checkmark       & \checkmark      & Simulation \\
\cline{1-2}\cline{4-8}\cline{10-14}
   \cite{jie2021dqn}                    & Single                       &                              & \LEFTcircle          & Homogeneous       & Edge                   & Homogeneous   & $\times$                           &                                                                                                  & DQN        & \checkmark & $\times$       & $\times$        & Simulation                   \\ 
\cline{1-2}\cline{4-8}\cline{10-14}
    \cite{9060882}                   & Multiple                     &                              & \LEFTcircle          & Homogeneous       & Edge                   & Homogeneous   & $\times$                           &                                                                                                  & DQN        & \checkmark           & $\times$       & $\times$        & Simulation                   \\ 
\cline{1-2}\cline{4-8}\cline{10-14}
    \cite{huang2019deep}                   & Multiple                     &                              & \LEFTcircle          & Heterogeneous     & Edge                   & Homogeneous   & $\times$                           &                                                                                                  & DQN        & \checkmark & $\times$       & \checkmark      & Simulation                   \\ 
\cline{1-2}\cline{4-8}\cline{10-14}
    \cite{8657791}                   & Single                       &                              & \LEFTcircle          & Homogeneous     & Edge                   & Homogeneous   & $\times$                           &                                                                                                  & DQN & \checkmark & \checkmark       & \checkmark      & Simulation                   \\ 
\cline{1-2}\cline{4-8}\cline{10-14}
    \cite{8493155}                   & Single                       &                              & \LEFTcircle          & Heterogeneous     & Edge                   & Homogeneous   & $\times$                           &                                                                                                  & Double DQN & \checkmark & $\times$       & \checkmark      & Simulation                   \\ 
\cline{1-2}\cline{4-8}\cline{10-14}
    \cite{9798187}                   & Single                       &                              & \LEFTcircle          & Homogeneous     & Edge                   & Homogeneous   & $\times$                           &                                                                                                  & SAC & \checkmark & $\times$       & $\times$      & Simulation                   \\
\cline{1-2}\cline{4-8}\cline{10-14}
    \cite{10136736}                   & Single                       &                              & \LEFTcircle          & Homogeneous     & Edge and Fog                   & Homogeneous   & $\times$                           &                                                                                                  & TD3 & \checkmark & $\times$       & \checkmark      & Simulation                   \\
\cline{1-2}\cline{4-8}\cline{10-14}
    \cite{li2022deep}                   & Single                       &                              & \Circle          & Homogeneous     & Edge                   & Homogeneous   & $\times$                           &                                                                                                  & DQN & $\times$ & \checkmark       & $\times$      & Simulation                   \\
\cline{1-2}\cline{4-8}\cline{10-14}
    \cite{liao2023online}                   & Single                       &                              & \LEFTcircle          & Homogeneous     & Edge                   & Homogeneous   & $\times$                           &                                                                                                  & DDPG and DQN & \checkmark & $\times$       & \checkmark      & Simulation                   \\
\cline{1-2}\cline{4-8}\cline{10-14}
    \cite{9945614}                   & Single                       &                              & \Circle          & Homogeneous     & Edge                   & Homogeneous   & $\times$                           &                                                                                                  & DDPG & \checkmark & $\times$       & \checkmark      & Simulation                   \\
\cline{1-2}\cline{4-8}\cline{10-14}
    \cite{sethi2023feddove}                   & Single                       &                              & \LEFTcircle          & Homogeneous     & Edge and Fog                   & Homogeneous   & $\times$                           &                                                                                                  & DQN & $\times$ & \checkmark       & \checkmark      & Simulation                   \\
 \cline{1-2}\cline{3-8}\cline{10-14}
\cite{xue2022deep}               & Multiple                     & \multirow{2}{*}{Dependent}                    & \LEFTcircle         & Heterogeneous     & Edge          & Heterogeneous & $\times$                         &                                                                                                  & GA and DQN        & \checkmark & $\times$     & $\times$      & Simulation                     \\   
\cline{1-2}\cline{4-8}\cline{10-14}
DRLIS               & Multiple                     &                     & \CIRCLE        & Heterogeneous     & Edge and Fog           & Heterogeneous & \checkmark                         &                                                                                                  & PPO        & \checkmark & \checkmark     & \checkmark      & Practical                    \\
\hline
\multicolumn{14}{l}{\CIRCLE: Real IoT Application and Deployment, \LEFTcircle: Simulated IoT Application, \Circle: Random} 
\end{tabular}%
}

\end{table*}

\renewcommand{\arraystretch}{1}

\section{System Model and Problem Formulation}
\label{system_model}
In this section, we first introduce the topology of the IoT systems in the edge and fog computing environment. Then, we discuss the problem formulation. The key notations are listed in Table~\ref{table: key_notations}.
\begin{table*}[]
\centering
\caption{List of key notations}
\label{table: key_notations}
\resizebox{\textwidth}{!}{%
\begin{tabular}{ll|ll}
\hline
\textbf{\textbf{Variable}}          & \textbf{\textbf{Description}}                                                                      & \textbf{\textbf{Variable}}             & \textbf{\textbf{Description}}                                                                                   \\ \hline
\textbf{$S$}                        & The application set                                                                                & \textbf{$\psi_{x_{S_{l_i}}}^{ram}$}    & The variance of RAM utilization of the server set after the scheduling configuration $x_{S_{l_i}}$              \\
\textbf{$S_l$}                      & One application (one task set)                                                                     & \textbf{$\Psi(\chi_{l})$}              & The load balancing model after the scheduling configuration $\chi_{l}$                                       \\
\textbf{$S_{l_i}$}                  & One task                                                                                           & \textbf{$\Psi(\chi)$}                  & The load balancing model after the scheduling configuration $\chi$                                              \\
\textbf{$N$}                        & The server set                                                                                     & \textbf{$\omega_{x_{S_{l_i}}}$}        & The total execution time (ms) for task $S_{l_i}$ based on the scheduling configuration $x_{S_{l_i}}$                 \\
\textbf{$x_{S_{l_i}}$}              & The scheduling configuration of task $S_{l_i}$                                                     & \textbf{$\omega_{x_{S_{l_i}}}^{trt}$}  & The ready time (ms) for task $S_{l_i}$ based on the scheduling configuration $x_{S_{l_i}}$             \\
\textbf{$\chi_{l}$}                 & The scheduling configuration of application $S_l$                                                  & \textbf{$\omega_{n_j,n_k}^{trt}$}      & The time (ms) consumed for required data by task $S_{l_i}$ to be sent from server $n_j$ to server $n_k$                    \\
\textbf{$\chi$}                     & The scheduling configuration of applications $S$                                                   & \textbf{$P(S_{l_i})$}                  & The parent tasks set of task $x_{S_{l_i}}$                                                                      \\
\textbf{$n_{k}^{cpu\_ut}$}          & The CPU utilization (\%) of server $n_k$                                                   & \textbf{$PS(S_{l_i})$}                 & The server set to which the dependency tasks of task $x_{S_{l_i}}$ are assigned                                 \\
\textbf{$n_{k}^{freq}$}             & The CPU frequency (MHz) of server $n_k$                                                      & \textbf{$\omega_{n_j,n_k}^{trans}$}    & The transmission time (ms) between server $n_j$ and server $n_k$                                       \\
\textbf{$n_{k}^{ram\_ut}$}          & The RAM utilization (\%) of server $n_k$                                                   & \textbf{$\omega_{n_j,n_k}^{prop}$}     & The propagation time (ms) between server $n_j$ and server $n_k$                                        \\
\textbf{$n_{k}^{ram\_size}$}        & The RAM size (GB) of server $n_k$                                                            & \textbf{$p_{n_j, n_k}$}                & The packet size (MB) from server $n_j$ to server $n_k$ for task $S_{l_i}$                                 \\
\textbf{${N}^{cpu\_uti}$}           & The CPU utilization (\%) of each server in server set $N$, denoted as a set                                                 & \textbf{$b_{n_j, n_k}$}                & The data rate (bit/s) between server $n_j$ and server $n_k$                                                     \\
\textbf{${N}^{ram\_uti}$}           & The RAM utilization (\%) of each server in server set $N$, denoted as a set                                                   & \textbf{$CP(S_{l_i})$}                 & Equals to $1$ if $S_{l_i}$ is on the critical path of application $S_l$, otherwise $0$                          \\
\textbf{$S_{l_i}^{ram}$}            & The minimum RAM required for executing task $S_{l_i}$                                              & \textbf{$\omega_{x_{S_{l_i}}}^{proc}$} & The processing time (ms) for task $S_{l_i}$ based on the scheduling configuration $x_{S_{l_i}}$        \\
\textbf{$\psi_{x_{S_{l_i}}}$}       & The load balancing model after the scheduling configuration $x_{S_{l_i}}$                          & \textbf{$\Omega(\chi_{l})$}            & The total execution time (ms) for application $S_l$ based on the scheduling configuration $\chi_{l}$   \\
\textbf{$\psi_{x_{S_{l_i}}}^{cpu}$} & The variance of CPU utilization of the server set after the scheduling configuration $x_{S_{l_i}}$ & \textbf{$\Omega(\chi)$}                & The total execution time (ms) for the application set $S$ based on the scheduling configuration $\chi$ \\ \hline
\end{tabular}%
}
\end{table*}

\subsection{System Model}
\begin{figure}
  \begin{center}
  \includegraphics[width=3.5in]{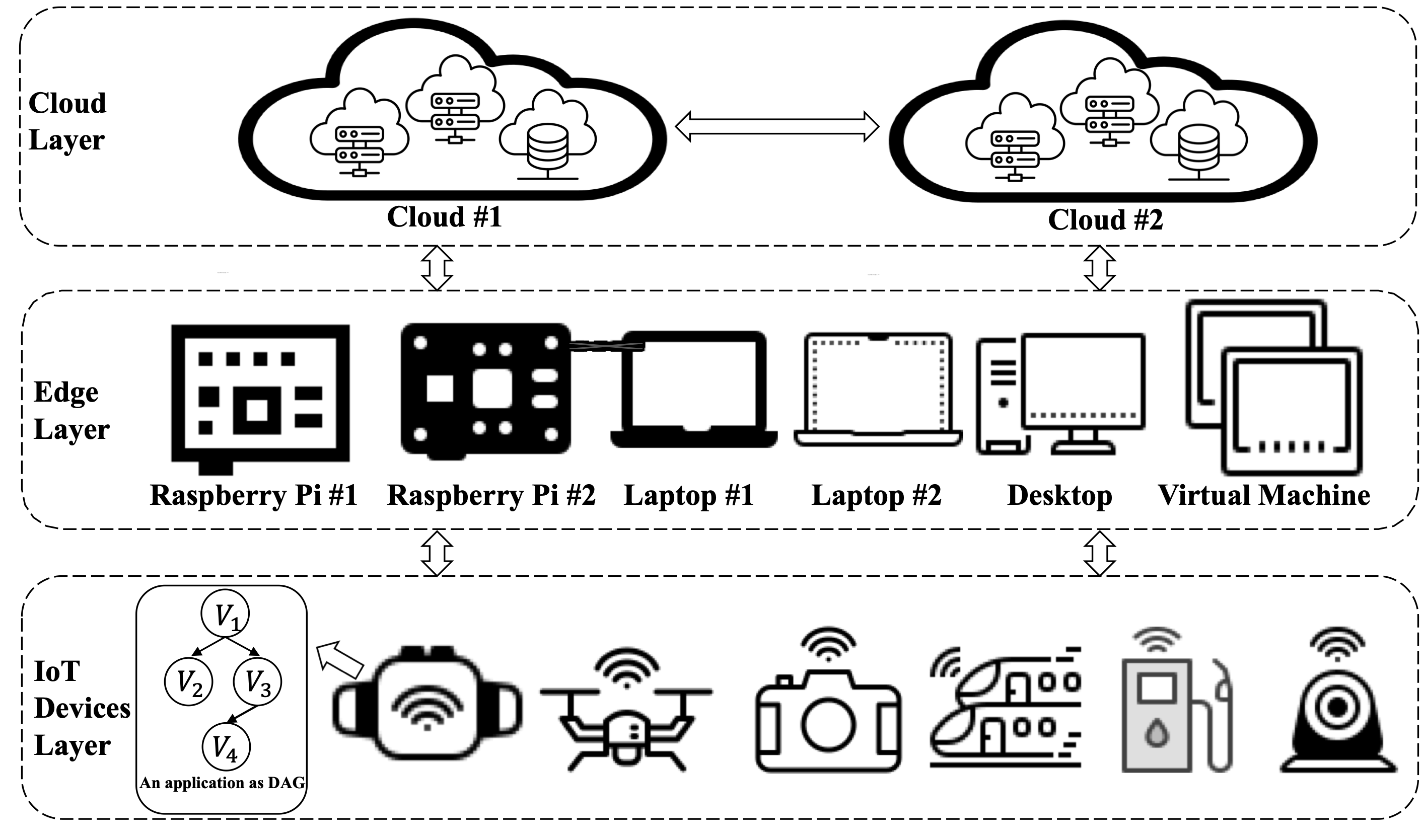}\\
  \caption{A view of the IoT system in fog computing}\label{iot_top}
  \end{center}
\end{figure}

\begin{figure}
  \begin{center}
  \includegraphics[width=4cm, height=5cm]{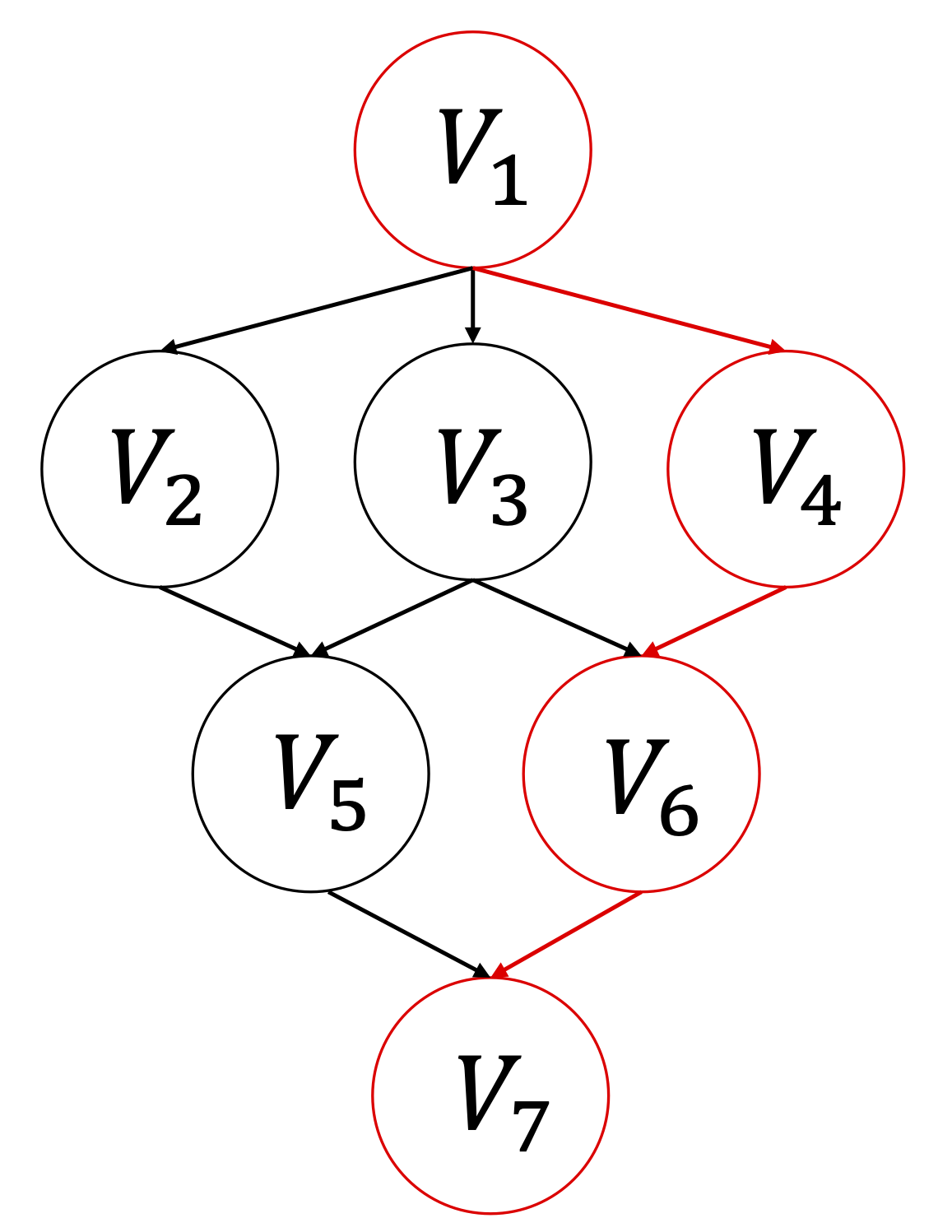}\\
  \caption{Sample IoT application with the critical path in red color}\label{dag}
  \end{center}
\end{figure}

Fig.~\ref{iot_top} represents a layered view of the IoT Systems in the fog computing environment. Consider $S = \{S_l|1 \leq l \leq |S|\}$ as a collection of $|S|$ applications, where each application contains one or more tasks, denoted as $S_l = \{S_{l_i}|1 \leq i \leq |S_l|\}$. The DAG $G = (V, E)$ is used to model an IoT application, as depicted in Fig.~\ref{dag}. A vertex $v_i = S_{l_i}$ denotes a certain task of the application, and an edge $e_{i,j}$ denotes the data flow between tasks $v_i$ and $v_j$, so some tasks must be executed after predecessor tasks are completed. $CP(S_l)$ represents the critical path (i.e., the path with the highest cost) of the DAG, marked in red in the figure.

\hspace{-0.5cm}

A set containing $|N|$ servers is used to process application set $S$, denoted as $N = \{n_k|1 \leq k \leq |N|\}$. To reflect the heterogeneity of the servers, for each server $n_k$, $n_{k}^{cpu\_ut}$ represents its CPU utilization (\%), $n_{k}^{freq}$ represents its CPU frequency (MHz), $n_{k}^{ram\_ut}$ represents its RAM utilization (\%), and $n_{k}^{ram\_size}$ represents its RAM size (GB). Moreover, $PS(S_{l_i})$ represents the server set to which the parent tasks of task $S_{l_i}$ are assigned, and $\omega_{n_j,n_k}^{trans}$, $\omega_{n_j,n_k}^{prop}$, {$p_{n_j, n_k}$}, and $b_{n_j, n_k}$ denote the transmission time (ms), the propagation time (ms), the packet size (MB), and the data rate (bit/s) between server $n_j$ and server $n_k$, respectively.

\subsection{Problem Formulation}
\label{problem_formulation}
Since an application contains one/multiple tasks, it may be executed on different servers. With a set of servers $N$, the scheduling configuration $x_{S_{l_i}}$ of a task $S_{l_i}$ is defined as:
\begin{equation}\label{eq:1}
x_{S_{l_i}} = \{n_k\},
\end{equation}
where $k$ shows the server's index. Accordingly, the scheduling configuration $\chi_{l}$ of an application $S_l$ is equal to the set of the scheduling configuration of the tasks it contains, defined as:
\begin{equation}
\chi_{l} = \{x_{S_{l_i}} | S_{l_i} \in S_l, 1 \leq i \leq |S_l| \}.
\end{equation}
The scheduling configuration $\chi $ of the application set $S$ is equal to the set of scheduling configuration per application:
\begin{equation}
\chi = \{\chi_{l} | 1 \leq l \leq |S| \}.
\end{equation}

In addition, we consider that for a given application, the execution model of tasks can be hybrid (i.e., sequential and/or parallel). That is, children tasks have some dependencies on the parent tasks that need to be executed after their completion, and we use $P(S_{l_i})$ to represent the parent task set of task $S_{l_i}$ \cite{zhu2023flight}. While tasks that do not depend on each other can be executed in parallel, and we use $CP(S_{l_i})$ to indicate that if a task $S_{l_i}$ is located on a critical path of application $S_l$. 

\subsubsection{\textbf{Load Balancing Model}}
\label{lb_model}
The load balancing model is used to measure the resource balancing level of the server set $N$ during the processing of the application set $S$. Regarding the server resource, both CPU and RAM are considered. For task $S_{l_i}$, the load balancing model $\psi_{x_{S_{l_i}}}$ is defined as:
\begin{equation}
\label{eq:4}
\psi_{x_{S_{l_i}}} = a_1\psi_{x_{S_{l_i}}}^{cpu} + a_2\psi_{x_{S_{l_i}}}^{ram},
\end{equation}
where $\psi_{x_{S_{l_i}}}^{cpu}$ and $\psi_{x_{S_{l_i}}}^{ram}$ represent the CPU and RAM models, and $a_1$ and $a_2$ are the control parameters by which the weighted load balancing model can be tuned. They satisfy:
\begin{equation}
a_1 + a_2 = 1,\; 0 \leq a_1,a_2 \leq 1.
\end{equation}
CPU model $\psi_{x_{S_{l_i}}}^{cpu}$ and RAM model $\psi_{x_{S_{l_i}}}^{ram}$ are defined as the variance of CPU and RAM utilization of the server set $N$ after the scheduling configuration $x_{S_{l_i}}$:
\begin{equation}
\psi_{x_{S_{l_i}}}^{cpu} = \text{Var}[{N}^{cpu\_uti}],
\end{equation}
\begin{equation}
\psi_{x_{S_{l_i}}}^{ram} = \text{Var}[{N}^{ram\_uti}],
\end{equation}
where
\begin{equation}
x_{S_{l_i}} = \{n_k\}. \tag{\ref{eq:1}}
\end{equation}

Correspondingly, for application $S_l$, the load balancing model $\Psi(\chi_{l})$ is defined as the sum of the load balancing models for each task processed by server set $N$:
\begin{equation}
\Psi(\chi_{l}) = \sum_{i=1}^{|S_l|}\psi_{x_{S_{l_i}}} \label{eq:9}.
\end{equation}

Our main goal is to find the best-possible scheduling configuration for the application set $S$ such that the variance of the overall CPU and RAM utilization of the server set $N$ during the processing of the application set $S$ can be minimized. Therefore, for the application set $S$, the load balancing model $\Psi(\chi)$ is defined as:
\begin{equation}
\Psi(\chi) = \sum_{l=1}^{|S|}\Psi(\chi_{l}) = \sum_{l=1}^{|S|}\sum_{i=1}^{|S_l|}\psi_{x_{S_{l_i}}}. \label{eq:10}
\end{equation}

\subsubsection{\textbf{Response Time Model}}
\label{rt_model}
We consider the response time model $\omega_{x_{S_{l_i}}}$ for the task $S_{l_i}$ consisting of two components, the task ready time model $\omega_{x_{S_{l_i}}}^{trt}$ and the processing model $\omega_{x_{S_{l_i}}}^{proc}$:
\begin{equation}
\label{eq:11}
\omega_{x_{S_{l_i}}} =  \omega_{x_{S_{l_i}}}^{trt} + \omega_{x_{S_{l_i}}}^{proc}. 
\end{equation}
The task ready time model $\omega_{x_{S_{l_i}}}^{trt}$ represents the maximum time for the data required by the task $S_{l_i}$ to arrive at the server to which it is assigned, defined as:
\begin{equation}
\omega_{x_{S_{l_i}}}^{trt} = max\;\;\omega_{n_j, n_k}^{trt}, \quad \forall n_j\in PS(S_{l_i}),
\end{equation}
where $\omega_{n_j, n_k}^{trt}$ denotes the time consumed for required data by task $S_{l_i}$ sent from server $n_j$ to server $n_k$, and $n_k$ is the server where the task $S_{l_i}$ will be executed based on scheduling configuration $x_{S_{l_i}}$, and $n_j$ represents the server where the parent task of task $S_{l_i}$ is executed. Therefore, $\omega_{n_j, n_k}^{trt}$ depends on the transmission time $\omega_{n_j, n_k}^{trans}$ and the propagation time $\omega_{n_j, n_k}^{prop}$ for task $S_{l_i}$ between server $n_j$ and server $n_k$:
\begin{equation}
\omega_{n_j, n_k}^{trt} = 
\begin{cases}
\omega_{n_j, n_k}^{trans} + \omega_{n_j, n_k}^{prop}& n_j \neq n_k,\\
0& n_j = n_k.
\end{cases}
\end{equation}
And the transmission time $\omega_{n_j, n_k}^{trans}$ can be calculated as:
\begin{equation}
\omega_{n_j, n_k}^{trans} = \frac{p_{n_j, n_k}}{b_{n_j, n_k}},
\end{equation}
where $p_{n_j, n_k}$ represents the packet size from server $n_j$ to server $n_k$ for task $S_{l_i}$, and $b_{n_j, n_k}$ represents the current bandwidth between server $n_j$ and server $n_k$ when the data for task $S_{l_i}$ is transmitted.

The processing model $\omega_{x_{S_{l_i}}}^{proc}$ is defined as the time it takes for assigned server $n_k$ to process the task $S_{l_i}$ based on scheduling configuration $x_{S_{l_i}}$, and can be calculated as:
\begin{equation}
\omega_{x_{S_{l_i}}}^{proc} = \frac{S_{l_i}^{size}}{n_{k}^{freq}}, 
\end{equation}
where $S_{l_i}^{size}$ represents the required CPU cycles for task $S_{l_i}$ and $n_{k}^{freq}$ represents the CPU frequency of server $n_k$ (for multi-core CPUs, the average frequency is considered).

Accordingly, the response time model $\Omega(\chi_{l})$ for application $S_l$ is defined as:
\begin{equation}
\Omega(\chi_{l}) = \sum_{i=1}^{|S_l|}(\omega_{x_{S_{l_i}}} \times CP(S_{l_i})), \label{eq:16}
\end{equation}
where $CP(S_{l_i})$ equals to $1$ if task $S_{l_i}$ is on the critical path of application $S_l$, otherwise $0$.

The main goal for the response time model $\Omega(\chi)$ is to find the best-possible scheduling configuration for the application set $S$ such that the total time for the server set $N$ processing them can be minimized. Therefore, for the application set $S$, the response time model $\Omega(\chi)$ is defined as:
\begin{equation}
\Omega(\chi) = \sum_{l=1}^{|S|}\Omega(\chi_{l}) = \sum_{l=1}^{|S|}\sum_{i=1}^{|S_l|}(\omega_{x_{S_{l_i}}} \times CP(S_{l_i})). \label{eq:17}
\end{equation}

\subsubsection{\textbf{Weighted Cost Model}}
\label{mo_model}
The weighted cost model is defined as the weighted sum of the normalized load balancing and normalized response time models. For task $S_{l_i}$:
\begin{equation}
\phi_{x_{S_{l_i}}} = w_1\frac{\psi_{x_{S_{l_i}}} - \psi^{min}}{\psi^{max} - \psi^{min}} + w_2\frac{\omega_{x_{S_{l_i}}} - \omega^{min}}{\omega^{max} - \omega^{min}},
\end{equation}
where $\psi_{x_{S_{l_i}}}$ and $\omega_{x_{S_{l_i}}}$ are the load balancing model and response time model of task $S_{l_i}$, and $\psi^{min}$, $\psi^{max}$, $\omega^{min}$, and $\omega^{max}$ represent the minimum and the maximum value of the load balancing model and response time model, respectively. Moreover, $w_1$ and $w_2$ are the control parameters by which the weighted cost model can be tuned. The reason we use the normalized models instead of the original models is that the values of the two models may be in different ranges. For example, the load balancing model may have a value from 0 to 1, while the response time model may have a value from 0 to 100. We need to normalize them so that the model values are in the same range.

Accordingly, the weighted cost model for application $S_l$ is defined as:
\begin{equation}
\Phi(\chi_{l}) = w_1 \times Norm(\Psi(\chi_{l})) + w_2 \times Norm(\Omega(\chi_{l})), 
\end{equation}
where $\Psi(\chi_{l})$ and $\Omega(\chi_{l})$ are obtained from Eq.~\ref{eq:9} and Eq.~\ref{eq:16}, and $Norm$ represents the normalization. The weighted cost model for the application set $S$ is defined as:
\begin{equation}
\label{eq:20}
\Phi(\chi) = w_1 \times Norm(\Psi(\chi)) + w_2 \times Norm(\Omega(\chi)), 
\end{equation}
where $\Psi(\chi)$ and $\Omega(\chi)$ are obtained from Eq.~\ref{eq:10} and Eq.~\ref{eq:17}.

Therefore, the weighted cost optimization problem of IoT applications can be formulated as:
\begin{align}
\label{eq:21}
     min \quad & \Phi(\chi) \\ 
  \text{s.t.} \quad 
     & C1:\;Size(x_{S_{l_i}}) = 1,\;\forall x_{S_{l_i}} \in \chi_{l} \\
     & C2:\;0 \leq n_k^{ram\_ut}, n_k^{cpu\_ut} \leq 1,\; \forall n_k \in N \\
     & C3:\;n_k^{freq}, n_k^{ram\_size} \geq 0,\; \forall n_k \in N \\
     & C4:\;S_{l_i}^{ram}<n_{k}^{ram\_size},\;\forall S_{l_i} \in S_l,\; \forall n_k \in N \\
     & C5:\;\Phi(x_{S_{l_j}}) \leq \Phi(x_{S_{l_j}} + x_{S_{l_i}}),\forall S_{l_j} \in P(S_{l_i}) \\
     & C6:\;w_1+w_2=1,\; 0 \leq w_1,w_2 \leq 1
\end{align}
where $C1$ states that any task can only be assigned to one server for processing. $C2$ states that for any server, the CPU utilization and RAM utilization are between $0$ and $1$. Besides, $C3$ states that the CPU frequency and the RAM size of any server are larger than $0$. Moreover, $C4$ denotes that any server should have sufficient RAM resources to process any task. Also, $C5$ denotes that any task can only be processed after its parent tasks have been processed, and thus the cumulative cost is always larger than or equal to the parent task. In addition, $C6$ denotes that the control parameters of the weighted cost model can only take value from $0$ to $1$, and the sum of them should be equal to $1$.

The problem being formulated is presented to be a non-convex optimization problem, because there may be an infinite number of local optima in the set of feasible domains, and usually, the complexity of the algorithm to find the global optimum is exponential (NP-hard) \cite{qiu2020distributed}. To cope with such non-convex optimization problems, most work decomposes them into several convex sub-problems and then solves these sub-problems iteratively until the algorithm converges \cite{tran2019federated}. This type of approach reduces the complexity of the original problem at the expense of accuracy \cite{ji2022trajectory}. In addition, such approaches are highly dependent on the current environment and cannot be applied in dynamic environments with complex and continuously changeable parameters and computational resources \cite{ji2022trajectory}. To deal with this problem, we propose DRLIS to efficiently handle uncertainties in dynamic environments by learning from interaction with the environment.

\section{Deep Reinforcement Learning Model}
\label{drl_model}
In reinforcement learning, the autonomous agent first interacts with the surrounding environment through action. Under the action and the environment, the agent generates a new state, while the environment gives an immediate reward. In this cycle, the agent interacts with the environment continuously and thus generates sufficient data. The reinforcement learning algorithm uses the generated data to modify its own action policy, then interacts with the environment to generate new data, and uses the new data to further improve its behavior. Formally, we use Markov Decision Process (MDP) to model the reinforcement learning problem. Specifically, the learning problem can be described by the tuple $<\mathbb{S},\mathbb{A},\mathbb{P},\mathbb{R},\gamma>$, where $\mathbb{S}$ denotes a finite set of states; $\mathbb{A}$ denotes a finite set of actions; $\mathbb{P}$ denotes the state transition probability; $\mathbb{R}$ denotes the reward function; $\gamma \in [0,1]$ is the discount factor, used to compute the cumulative rewards.

We assume that the time $\mathbb{T}$ of the learning process is divided into multiple time steps $t$ and the agent will interact with the environment at each time step and have multiple states $S_t$. At a particular time step $t$, the agent possesses the environment state $S_t = s$, where $s \in \mathbb{S}$. The agent chooses an action $A_t = a$ according to the policy $\pi(a|s)$, where $a \in \mathbb{A}$, and $\pi(a|s) = Pr[A_t=a|S_t=s]$ is the policy function, which denotes the probability of choosing the action $a$ in state $s$. After choosing action $a$, the agent receives a reward $r=\mathbb{R}[S_t = s, A_t = a]$ from the environment based on the reward function $\mathbb{R}$, and it moves to the next state $S_{t+1} = s'$ based on the state transition function $P^a_{ss'} = \mathbb{P}[S_{t+1} = s'|S_t = s, A_t = a]$. The goal of the reinforcement learning agent is to learn a policy $\pi$ that maximizes the expectation of cumulative discounted reward $\mathbb{E}_\pi[\sum_{t \in T}\gamma_tr_t]$.

Based on the weighted cost optimization problem of IoT applications in edge and fog computing environments, the state space $\mathbb{S}$, action space $\mathbb{A}$, and reward function $\mathbb{R}$ for the MDP are defined as follows:
\begin{itemize}
\item \textbf{State space $\mathbb{S}$}: Since the optimization problem is related to tasks and servers, the state of the problem consists of the feature space of the task currently being processed and the state space of the current server set $N$. Based on the discussion in Section~\ref{system_model}, at the time step $t$, the feature space of the task $S_{l_i}$ includes the task ID, the tasks' predecessors and successors, the application ID to which the task belongs, the number of tasks in the current application, the estimate of the occupied CPU resources for the execution of the task, the task's RAM requirements, the estimate of the task's response time, etc. Formally, the feature space $\mathbb{F}$ for task $S_{l_i}$ at the time step $t$ is defined as follows:
\begin{equation}
 \mathbb{F}_t(S_{l_i}) = \{f_t^y(S_{l_i})| S_{l_i} \in S_l, 0 \leq y \leq |\mathbb{F}|\},
\end{equation}
where $y$ represents the index of the feature in the task feature space $\mathbb{F}$, and $|\mathbb{F}|$ represents the number of features. Moreover, at the time step $t$, the state space of the current server set $N$ includes the number of servers, each server's CPU utilization, CPU frequency, RAM utilization, and RAM size, and the propagation time and bandwidth between different servers, etc. Formally, the state space 
$\mathbb{G}$ for the server set $N$ at the time step $t$ is defined as:
\begin{equation}
\begin{split}
 \mathbb{G}_t(N) = \{|N|, g_t^z(n_k), h_t^q(n_j, n_k)| n_j, n_k \in N, \\
 0 \leq z \leq |g|, 0 \leq q \leq |h|\},
 \end{split}
\end{equation}
where $g$ represents the state type that is related to only one server (i.e., CPU utilization), $z$ represents its index, and $|g|$ represents the length of this type of state; besides, $h$ denotes the state type that is related to two servers (i.e., propagation time), and similarly, $q$ represents its index and $|h|$ represents the length of this type of state. Therefore, the state space $\mathbb{S}$ is defined as:
\begin{equation}
 \mathbb{S} = \{S_t = (\mathbb{F}_t(S_{l_i}), \mathbb{G}_t(N))|S_{l_i} \in S_l, t \in \mathbb{T}\}.
\end{equation}
\item \textbf{Action space $\mathbb{A}$}: The goal is to find the best-possible scheduling configuration for the application set $S$ to minimize the objective function Eq.~\ref{eq:21}. Therefore, at the time step $t$, the action can be defined as the assignment of the server to the task $S_{l_i}$:
\begin{equation}
A_t = x_{S_{l_i}} = n_k.
\end{equation}
Accordingly, the action space $\mathbb{A}$ can be defined as the server set $N$:
\begin{equation}
\mathbb{A} = N.
\end{equation}
\item \textbf{Reward function $\mathbb{R}$}: Since this is a weighted cost optimization problem, we need to define the reward function for each sub-problem. First, as the $penalty$, a very large negative value is introduced if the task cannot be processed on the assigned server for any reason. Also, for the load balancing problem, based on the discussion in section~\ref{lb_model}, the reward function $r_t^{lb}$ is defined as:
\begin{equation}
\label{eq:33}
r_t^{lb} = 
\begin{cases}
\psi_{x_{S_{l_{i-1}}}} - \psi_{x_{S_{l_i}}}& succeed \\
penalty& fail,
\end{cases}
\end{equation}
where $\psi_{x_{S_{l_i}}}$ is obtained from Eq.~\ref{eq:4}. The value output by reward function $r_t^{lb}$ is the difference between the load balancing models of the server set after scheduling the current task and the previous one. If the value of the load balancing model of the server set is reduced after scheduling the current task, the output reward is positive, otherwise it is negative. Beside, for the response time problem, based on the discussion in section \ref{rt_model}, the reward function $r_t^{rt}$ is defined as:
\begin{equation}
\label{eq:34}
r_t^{rt} = 
\begin{cases}
\omega_{x_{S_{l_i}}}^{mean} - \omega_{x_{S_{l_i}}}& succeed \\
penalty& fail,
\end{cases}
\end{equation}
where $\omega_{x_{S_{l_i}}}$ is obtained from Eq. \ref{eq:11}, and $\omega_{x_{S_{l_i}}}^{mean}$ represents the average response time for task $S_{l_i}$. The value output by reward function $r_t^{rt}$ is the difference between the average response time (the current response time is also considered) and the current response time for task $S_{l_i}$. If the current response time is lower than the average one, the output reward is positive, otherwise it is negative. The reward function $r_t$ for the weighted cost optimization problem is defined as:
{
\small
\begin{equation}
\hspace{-1cm}
\label{eq:35}
r_t =
\begin{cases}
w_1 \times Norm(r_t^{lb}) + w_2 \times Norm(r_t^{rt})& succeed \\
penalty& fail,
\end{cases}
\end{equation}}
where $w_1$ and $w_2$ are the control parameters, and $Norm$ represents the normalization process.
\end{itemize}

Currently, many advanced deep reinforcement learning algorithms (e.g., PPO, TD3, SAC) have been proposed by different researchers. They show excellent performance in different fields. PPO improves convergence and sampling efficiency by adopting importance sampling and proportional clipping \cite{schulman2017proximal}. TD3 (Twin Delayed DDPG) introduces a dual Q network and delayed update strategy to effectively solve the overestimation problem in the continuous action space \cite{pmlr-v80-fujimoto18a}. SAC (Soft Actor-Critic) combines policy optimization and learning of Q-value functions, providing more robust and exploratory policy learning through maximum entropy theory \cite{pmlr-v80-haarnoja18b}. These algorithms have achieved remarkable results in different tasks and environments. In our research problem, the agent's action and state space is discrete, which hinders the application of TD3, because it is designed for continuous control \cite{9619008}. In addition, the original SAC only considers the problem of continuous space \cite{pmlr-v80-haarnoja18b}, although there are some works discussing how to apply SAC to discrete space, they usually need to adopt some special tricks and extensions, such as using soft-max or sample-prune techniques to accommodate discrete actions \cite{christodoulou2019soft}. Besides, Wang et al. \cite{wang2023comparative} shows that SAC requires more computation time and convergence time than PPO. Whereas our study focuses on edge and fog computing environments, where handling latency sensitivity and variation are important considerations for choosing the appropriate DRL algorithm. We choose PPO as the basis of DRLIS, because PPO is designed to be more easily adaptable to discrete action spaces \cite{zhu2021overview} and we aim for the algorithm to converge quickly and perform well in diverse environments.

\section{DRL-based Optimization Algorithm}
\label{ppo_algorithm}
Based on the above-mentioned MDP model, we propose DRLIS to achieve weighted cost optimization of IoT applications in edge and fog computing environments. In this section, we introduce the mathematical principle of the PPO algorithm and discuss the proposed DRLIS.

\subsection{Preliminaries}
\label{preliminaries}
The PPO algorithm belongs to the Policy Gradient (PG) algorithm which considers the impact of actions on rewards and adjusts the probability of actions \cite{sutton1999policy}. We use the same notations as in section \ref{system_model} to describe the algorithm. We consider the time horizon $\mathbb{T}$ is divided into multiple time steps $t$, and the agent has a policy $\pi_\theta$ for determining its actions and interactions with the environment. The objective can be expressed as adjusting the parameter $\theta$ to maximize the expected cumulative discounted rewards $\mathbb{E}_{\pi_\theta}[\sum_{t \in T}\gamma_tr_t]$ \cite{schulman2017proximal}, expressed by the formula:
\begin{equation}
J(\theta) = \mathbb{E}_{\pi_\theta}[\sum_{t \in T}\gamma_tr_t].
\end{equation}
Since this is a maximization problem, the gradient ascent algorithm can be used to find the maximum value:
\begin{equation}
\theta' = \theta + \alpha \nabla_\theta J(\theta).
\end{equation}
The key is to obtain the gradient of the reward function $J(\theta)$ with respect to $\theta$, which is called the policy gradient. The algorithm for solving reinforcement problems by optimizing the policy gradient is called the policy gradient algorithm. The policy gradient can be presented as,
\begin{equation}
\nabla_\theta J(\theta) = \mathbb{E}_{\pi_\theta}[\nabla_\theta log \pi_\theta(a_t|s_t) A_\theta(a_t|s_t)],
\end{equation}
where $A_\theta(a_t|s_t)$ is the advantage function at time step t, used to evaluate the action $a_t$ at the state $s_t$. Here, the policy gradient indicates the expectation of $\nabla_\theta log \pi_\theta(a_t|s_t) A_\theta(a_t|s_t)$, which can be estimated using the empirical average obtained by sampling. However, the PG algorithm is very sensitive to the update step size, and choosing a suitable step size is challenging \cite{huang2020policy}. Moreover, practice shows that the difference between old and new policies in training is usually large \cite{schulman2017proximal}. 

To address this problem, Trust Region Policy Optimization (TRPO) \cite{schulman2015trust} is proposed. This algorithm introduces importance sampling to evaluate the difference between the old and new policies and restricts the new policy if the importance sampling ratio grows large. Importance sampling refers to replacing the original sampling distribution with a new one to make sampling easier or more efficient. Specifically, TRPO maintains two policies, the first policy $\pi_{\theta_{old}}$ is the current policy to be refined, and the second policy $\pi_\theta$ is used to collect the samples. The optimization problem is defined as follows:
\begin{align}
\label{eq:39}
\underset{\theta}{\text{maximize}} \quad & \mathbb{E}_t[\frac{\pi_\theta(a_t|s_t)}{\pi_{\theta_{old}}(a_t|s_t)}A_t] \\ 
\text{subject to} \quad & \mathbb{E}_t[KL[\pi_{\theta_{old}}(\cdot|s_t), \pi_\theta(\cdot|s_t)]] \leq \delta,
\end{align}
where $KL$ represents Kullback-Leibler Divergence, used to quantify the difference between two probability distributions \cite{van2014renyi}, and $\delta$ represents the restriction of the update between old policy $\pi_{\theta_{old}}$ and new policy $\pi_\theta$. After linear approximation of the objective and quadratic approximation of the constraints, the problem can be efficiently approximated using the conjugate gradient algorithm. However, the computation of conjugate gradient makes the implementation of TRPO more complex and inflexible in practice \cite{shao2017customised}, \cite{li2019hierarchical}.

To make this algorithm well applied in practice, the KL-PPO algorithm \cite{schulman2017proximal} is proposed. Rather than using the constraint function $\mathbb{E}_t[KL[\pi_{\theta_{old}}(\cdot|s_t), \pi_\theta(\cdot|s_t)]] \leq \delta$, the $KL$ divergence is added as a penalty in the objective function:
{
\small
\begin{equation}
\hspace{-0.5cm}
L^{KLPEN}(\theta) = \mathbb{E}_t[r_t(\theta)A_t - \beta KL[\pi_{\theta_{old}}(\cdot|s_t), \pi_\theta(\cdot|s_t)]],
\end{equation}
}
where $r_t(\theta) = \frac{\pi_\theta(a_t|s_t)}{\pi_{\theta_{old}}(a_t|s_t)}$ is the ratio of the new policy and the old policy, obtained in Eq. \ref{eq:39}, and the parameter $\beta$ can be dynamically adjusted during the iterative process according to the $KL$ divergence. If the current $KL$ divergence is larger than the predefined maximum value, indicating that the penalty is not strong enough and the parameter $\beta$ needs to be increased. Conversely, if the current $KL$ divergence is smaller than the predefined minimum value, the parameter $\beta$ needs to be reduced.

Moreover, another idea to restrict the difference between old policy $\pi_{\theta_{old}}$ and new policy $\pi_\theta$ is to use clipped surrogate function $clip$. The PPO algorithm using the clip function (CLIP-PPO) removes the KL penalty and the need for adaptive updates to simplify the algorithm. Practice shows CLIP-PPO usually performs better than KL-PPO \cite{schulman2017proximal}. Formally, the objective function of CLIP-PPO is defined as follows:
{
\small
\begin{equation}
\hspace{-0.5cm}
\label{eq:42}
L^{CLIP}(\theta) = \mathbb{E}_t[min(r_t(\theta)A_t, clip(r_t(\theta), 1-\epsilon, 1+\epsilon)A_t].
\end{equation}
}
And $clip(r_t(\theta), 1-\epsilon, 1+\epsilon)$ restrict the ratio $r_t(\theta)$ into $(1 - \epsilon, 1 + \epsilon)$, defined as:
{
\small
\begin{equation}
\hspace{-0.5cm}
clip(r_t(\theta), 1-\epsilon, 1+\epsilon) = 
\begin{cases}
1-\epsilon& r_t(\theta) < 1-\epsilon \\
r_t(\theta)& 1-\epsilon \leq r_t(\theta) \leq 1+\epsilon \\
1+\epsilon& r_t(\theta) > 1+\epsilon. \\
\end{cases}
\end{equation}
}

By removing the constraint function as discussed in TRPO, both PPO algorithms significantly reduce the computational complexity, while ensuring that the updated policy deviates not too large from the previous one.

\subsection{DRLIS: DRL-based IoT Application Scheduling}
Since CLIP-PPO usually outperforms KL-PPO in practice, we choose it as the basis for the optimization algorithm. DRLIS is based on the actor-critic framework, which is a reinforcement learning method combining Policy Gradient and Temporal Differential (TD) learning. As the name implies, this framework consists of two parts, the actor and the critic, and in implementation, they are usually presented as Deep Neural Networks (DNNs). The actor network is used to learn a policy function $\pi_\theta(a|s)$ to maximize the expected cumulative discounted reward $\mathbb{E}_\pi[\sum_{t \in T}\gamma_tr_t]$, while the critic network is used to evaluate the current policy and to guide the next stage of the actor's action. In the learning process, at the time step $t$, the reinforcement learning agent inputs the current state $s_t$ into the actor network, and the actor network outputs the action $a_t$ to be performed by the agent in the MDP. The agent performs the action $a_t$, receives the reward $r_t$ from the environment, and moves to the next state $s_{t+1}$. The critic network receives the states $s_t$ and $s_{t+1}$ as input and estimates their value functions $V_{\pi_\theta}(s_t)$ and $V_{\pi_\theta}(s_{t+1})$. The agent then computes the TD error $\delta_t$ for the time step t:
\begin{equation}
\delta_t = r_t + \gamma V_{\pi_\theta}(s_{t+1}) - V_{\pi_\theta}(s_t),
\end{equation}
where $\gamma$ denotes the discount factor, as discussed in section \ref{system_model}, and the actor network and critic network update their parameters using the TD error $\delta_t$. DRLIS continues this process after multiple steps, as an estimate $\hat{A}_t$ of the advantage function $A_t$, which can be written as:
{
\footnotesize
\begin{equation}
\label{eq:45}
\hat{A}_t = - V_{\pi_\theta}(s_t) + r_t + \gamma r_{t+1} + \cdot\cdot\cdot + \gamma^{T-t+1} r_{T-1} + \gamma^{T-t}V_{\pi_\theta}(s_T).
\end{equation}
}
DRLIS maintains three networks, one critical network, and two actor networks (i.e., the old actor and the new actor), representing the old policy function $\pi_{\theta_{old}}$ and the new policy function $\pi_\theta$, as discussed in section \ref{preliminaries}. Algorithm \ref{ppo} describes DRLIS for the weighted cost optimization problem in edge and fog computing environments.

\begin{algorithm}[h]
\footnotesize
\caption{DRLIS for weighted cost optimization}\label{ppo}
\SetKwInOut{Input}{Input}
\SetKwInOut{Initialization}{Initialization}
\Input{new actor network $\Pi_\theta$ with parameter $\theta$; old actor network $\Pi_{\theta_{old}}$ with parameter $\theta_{old}$, where $\theta_{old}=\theta$; critic network $V_\mu$ with parameter $\mu$; max time step $T$; update epoch $K$;  policy objective function coefficient $a_c$; value function loss function coefficient $a_v$; entropy bonus coefficient $a_e$; clipping ratio $\epsilon$} 
\While{True}{
    $servers \gets GetServers()$\;
    $task \gets GetTask()$\;
    \If{$servers \neq  servers_{old}$}{
        $agent \gets InitializeAgent(servers)$\;
        $servers_{old} \gets servers$\;
    }
    $s_1 \gets GeneralizeState(servers, task)$\;
    $\mathcal{D} \gets InitializeBuffer()$\;
    \For{$t \gets 1$ to $T$}{
        $a_t \gets \Pi_\theta(s_t)$\;
        $Schedule(task, a_t)$\;
        $r_t \gets GetReward()$\;
        $servers \gets GetServers()$\;
        \If{$servers \neq  servers_{old}$}{
            $break$\;
        }
        $task \gets GetTask()$\;
        $s_{t+1} \gets GeneralizeState(servers, task)$\;
        $u_t = (s_t, a_t, r_t)$\;
        $\mathcal{D}.Append(u_t)$\;
    }
     $\hat{A}_t \gets - V_\mu(s_t) + r_t + \gamma r_{t+1} + \cdot\cdot\cdot + \gamma^{T-t+1} r_{T-1} + \gamma^{T-t}V_\mu(s_T)$\;
    \For{$k \gets 1$ to $K$}{
        $L^{CLIP}(\theta) = \sum_1^t min(\frac{\Pi_\theta(a_t|s_t)}{\Pi_{\theta_{old}}(a_t|s_t)}\hat{A}_t, clip(\frac{\Pi_\theta(a_t|s_t)}{\Pi_{\theta_{old}}(a_t|s_t)}, 1-\epsilon, 1+\epsilon)\hat{A}_t$\;
        $L^{VF}(\mu) = \sum_1^t(V_\mu(s_t) - \hat{A}_t)^2$\;
        $L^{ET}(\theta) = \sum_1^t Entropy(\Pi_\theta(a_t|s_t))$\;
        $L(\theta,\mu) = - a_c L^{CLIP}(\theta) + a_v L^{VF}(\mu) - a_e L^{ET}(\theta)$\;
        update $\theta$ and $\mu$ with $L(\theta,\mu)$ by Adam optimizer\;
    }
    $\theta_{old} \gets \theta$\;
}
\end{algorithm}
We consider a scheduler that is implemented based on DRLIS. When this scheduler receives a scheduling request from an IoT application, it obtains information about the set of servers currently available and initializes a DRL agent based on the information. This agent contains three deep neural networks, a new actor network $\Pi_\theta$ with parameter $\theta$, an old actor network $\Pi_{\theta_{old}}$ with parameter $\theta_{old}$, where $\theta_{old}=\theta$, and a critic network $V_\mu$ with parameter $\mu$. After that, the scheduler obtains the information about the currently submitted task and generates the current state $s_t$ based on the information regarding the task and servers. Inputting the state $s_t$ to the new actor network $\Pi_\theta$ will output an action $a_t$, representing the target server to which the current task is to be assigned. The scheduler then assigns the task to the target server and receives the corresponding reward $r_t$, which is calculated based on Eq. \ref{eq:33}, \ref{eq:34}, \ref{eq:35}. The reward $r_t$ is essential for indicating the positive or negative impact of the agent's current scheduling policy on the optimization objectives (e.g., IoT application response time and servers load balancing level). Also, a tuple $u_t$ with three values $(s_t, a_t, r_t)$ will be stored in buffer $\mathcal{D}$. The scheduler repeats the process $T$ times until sufficient information is collected to update the neural networks. When updating the neural networks, the estimate of the advantage function is first computed based on Eq. \ref{eq:45}. Then the neural networks are optimized for K times. Both actor network and critic network use Adam optimizer, and the loss function is computed as:
\begin{equation}
L(\theta,\mu) =  - a_c L^{CLIP}(\theta) + a_v L^{VF}(\mu) - a_e L^{ET}(\theta),
\end{equation}
where $L^{CLIP}(\theta)$ is the policy objective function from Eq. \ref{eq:42}, and $L^{VF}(\mu)$ is loss function for the state value function: 
\begin{equation}
L^{VF}(\mu) = \sum_{t=1}^T(V_\mu(s_t) - \hat{A}_t)^2.
\end{equation}
And $L^{ET}(\theta)$ is the entropy bonus for the current policy:
\begin{equation}
L^{ET}(\theta) = \sum_{t=1}^T Entropy(\Pi_\theta(a_t|s_t)).
\end{equation}
In addition, $a_c$, $a_v$, and $a_e$ are the coefficients. After updating the neural networks, the parameter $\theta$ of the new actor network $\Pi_\theta$ will be copied to the old actor network $\Pi_{\theta_{old}}$. Assuming that there are $N$ tasks, from Algorithm \ref{ppo}, the agent will update the policy K times after scheduling T tasks, so the complexity of the algorithm as $O(N+\frac{N}{T}K)$. In practical applications, both $T$ and $K$ as hyperparameters can be customized to suit different computational environments. Thus the computational complexity of the algorithm actually depends on the number of tasks $N$ and can be written as $O(N)$. For the edge/fog environment with limited computational resources, we consider this computational complexity to be acceptable.

\subsection{Practical Implementation in the FogBus2 Framework}
\label{implementation}
We extend the scheduling module of the FogBus2 framework\footnote{https://github.com/Cloudslab/FogBus2} \cite{deng2021fogbus2} to design and develop the DRLIS in practice for processing placement requests from different IoT applications in edge and fog computing environments. 

\begin{figure}[!htb]
\centering
\includegraphics[width=\linewidth]{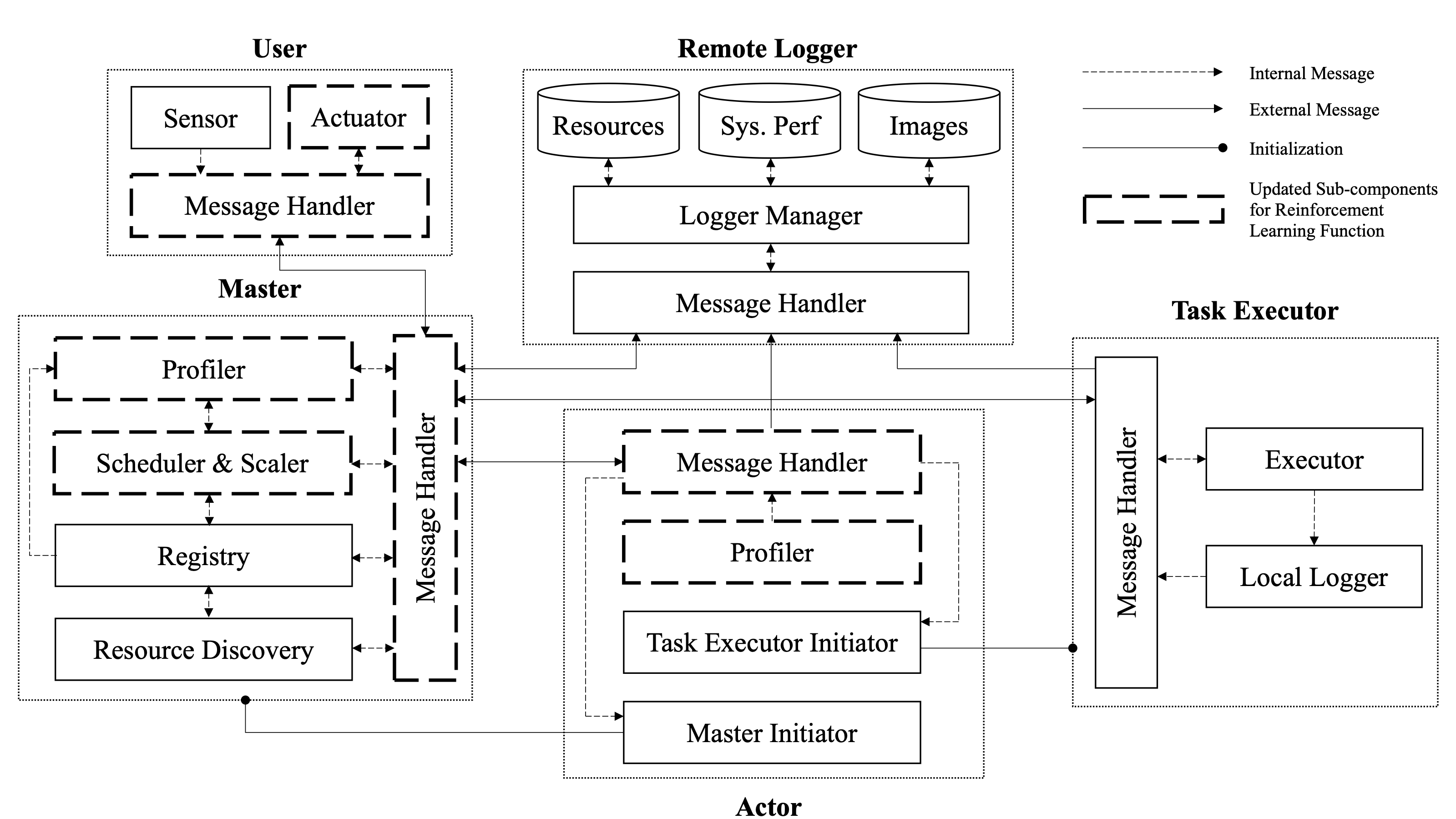}
\caption{Updated Sub-Components for Reinforcement Learning in FogBus2 Framework}
\label{fig:fgbs}
\end{figure}

FogBus2 is a lightweight container-based distributed/ serverless framework (realized using Docker microservices software) for integrating edge and fog/cloud computing environments. A scheduling module is implemented to decide the deployment of heterogeneous IoT applications, enabling the management of distributed resources in the hybrid computing environment. There are five main components within FogBus2 framework, namely \textit{Master}, \textit{Actor}, \textit{RemoteLogger}, \textit{TaskExecutor}, and \textit{User}. Fig. \ref{fig:fgbs} shows the relationship between different components in the FogBus2 framework, and the updated sub-components used to implement the reinforcement learning function.

\begin{itemize}
\item \textit{Remote Logger}: It is designed for collecting and storing logs from other components, whether periodic or event-driven.
\item \textit{Master}: It contains the scheduling module of FogBus2, responsible for the registration and scheduling of IoT applications. It can also discover resources and self-scale based on the input load. We implement a reinforcement learning scheduling module in the \textit{Scheduler \& Scaler} sub-component. Besides, we extend the functionality of the \textit{Profiler} and the \textit{Message Handler} components to allow \textit{Master} components to receive and handle information from other components for reinforcement learning scheduling.
\item \textit{Actor}: It informs the \textit{Remote Logger} and \textit{Master} components of the computing resources of the corresponding node to coordinate the resource scheduling of the framework. Furthermore, it is responsible for launching the appropriate \textit{Task Executor} components to process the submitted IoT application. We extend the functionality of the \textit{Profiler} and the \textit{Message Handler} components to allow system characteristics regarding servers to be passed to the reinforcement learning scheduling module in \textit{Master} components.
\item \textit{Task Executor}: It is responsible for executing the corresponding tasks of the submitted application. The results are passed to the \textit{Master} component.
\item \textit{User}: It runs on IoT devices and is responsible for processing raw data from sensors and users. It sends the processed data to the \textit{Master} component and submits the execution request. We extend the functionality of the \textit{Actuator} and the \textit{Message Handler} components to allow information related to IoT applications to be passed to the reinforcement learning scheduling module in \textit{Master} components.
\end{itemize}

\begin{figure}[!htb]
\centering
\includegraphics[width=\linewidth]{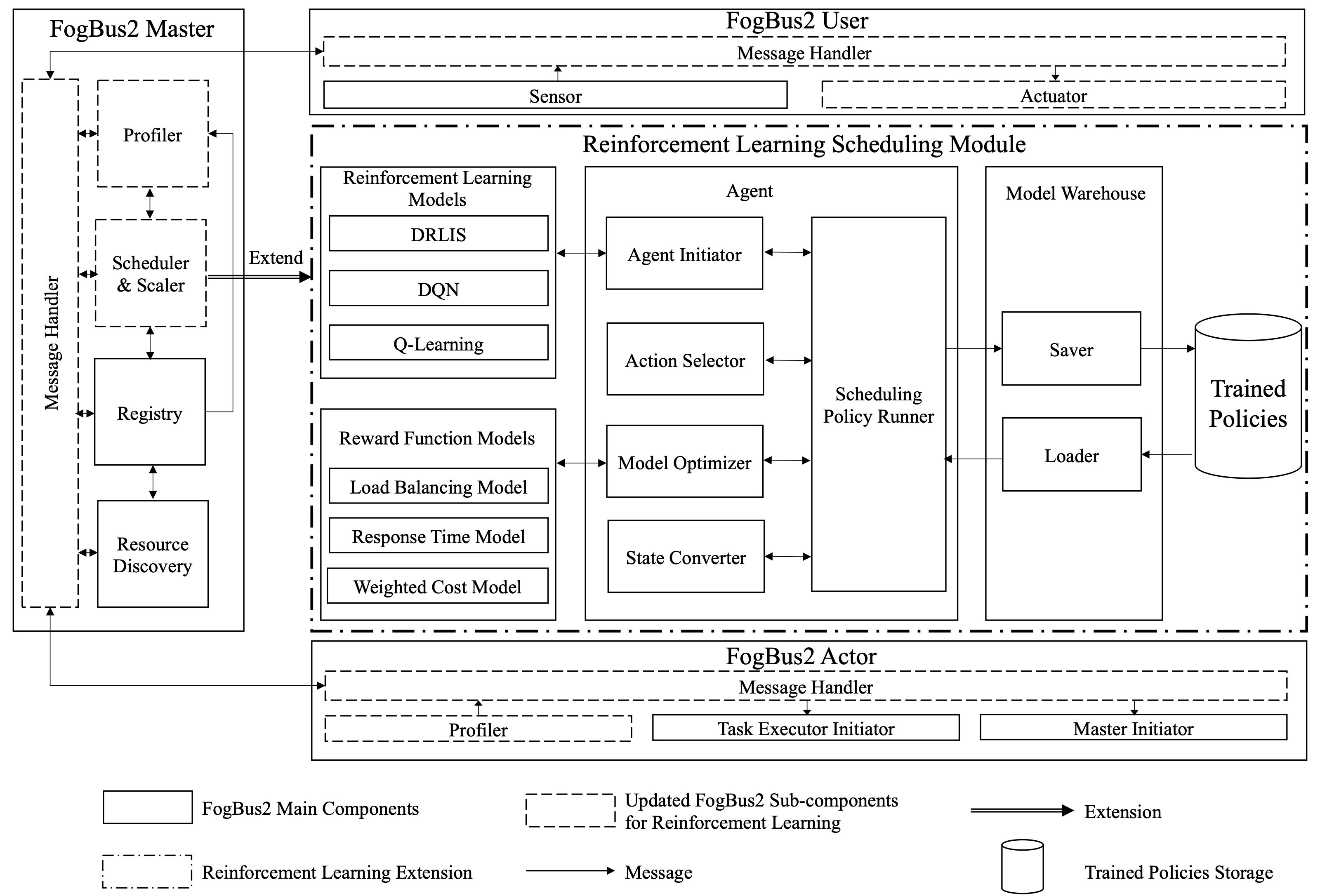}
\caption{Reinforcement Learning Scheduling Module in FogBus2 Framework}
\label{fig:fgbs_rf}
\end{figure}

Fig. \ref{fig:fgbs_rf} shows our implementation of the reinforcement learning scheduling module in the FogBus2 framework. The module can be divided into four sub-modules: 1) Reinforcement Learning Models, 2) Rewards Models, 3) Reinforcement Learning Agent, and 4) Model Warehouse.

\begin{itemize}
\item \textit{Reinforcement Learning Models}: This sub-module contains the reinforcement learning models. According to Algorithm \ref{ppo}, we implement a DRLIS-based model. In addition, to evaluate the performance of DRLIS, we also implement DQN and Q-Learning-based models. 
\item \textit{Rewards Models}: This sub-module contains the models associated with the reward functions. According to Section \ref{problem_formulation} and Section \ref{drl_model}, we implemented Load Balancing Model, Response Time Model, and Weighted Cost Model. This sub-module is responsible for calculating the reward values based on the information (e.g., CPU and RAM utilization) and transferring them to the \textit{Agent} sub-module.
\item \textit{Reinforcement Learning Agent}: This sub-module implements the functions of the reinforcement learning agent. The \textit{Agent Initiator} calls the \textit{Reinforcement Learning Models} sub-module and initializes the corresponding models. The \textit{Action Selector} is responsible for outputting the target server index for the currently scheduled task. The \textit{Model Optimizer} optimizes the running reinforcement learning scheduling policy based on the reward values returned from the \textit{Reward Function Models} sub-module. The \textit{State Converter} is responsible for converting the parameters of the server and IoT application into state vectors that can be recognized by the reinforcement learning scheduling model. The \textit{Scheduling Policy Runner} is the running program of the reinforcement learning scheduling \textit{Agent} and is responsible for receiving submitted tasks, saving or loading the trained policies, and requesting and accessing parameters from other FogBus2 components (e.g., \textit{FogBus2 Actor}, \textit{FogBus2 User}) for the computation of reward functions.
\item \textit{Model Warehouse}: This sub-module can save the hyperparameters of the trained scheduling policy to the database and loads the hyperparameters to initialize a well-trained scheduling \textit{Agent}.
\end{itemize}

Algorithm \ref{scheduler} summarizes the scheduling mechanism based on DRLIS.
\begin{algorithm}[h]
\footnotesize
\caption{Reinforcement learning scheduler in FogBus2 framework based on the proposed weighted cost optimization algorithm}\label{scheduler}
\SetKwInOut{Input}{Input}
\SetKwInOut{Initialization}{Initialization}
\Input{master component $M$; registered actor component set $A$; user component $U$; tasks to be processed $T$} 
$Scheduler \gets InitializeScheduer(DRLIS)$\;
$\mathcal{D}_A \gets InitializeActorBuffer()$\;
$\mathcal{D}_U \gets InitializeUserBuffer()$\;
\While{True}{
    $U.SubmitTasks(T)$\;
    $AvailableActors \gets M.CheckResources(T)$\;
    \If{$AvailableActors$ is $empty$}{
        $M.Message(U,Fail)$\;
        $break$\;
    }
    \ForEach{$t_i \in T$}{
        $Scheduler.TaskPlacement(t_i, A)$\;
        $A.Message(M, I_A)$\;
        $\mathcal{D}_A.Append(I_A)$\;
        $U.Message(M, I_U)$\;
        $\mathcal{D}_U.Append(I_U)$\;
        \If{$UpdateScheduler$ is $True$}{
            $Rewards \gets ComputeRewards(\mathcal{D}_A, \mathcal{D}_U)$\;
            $Scheduler.Update()$\;
        }    
    }
}
\end{algorithm}
The framework first initializes a scheduler, based on Algorithm \ref{ppo}. In addition, two buffers $\mathcal{D}_A$ and $\mathcal{D}_U$ for storing information from the $Actor$ component and the $User$ component are also initialized. After the $User$ component submits the IoT application to be processed, the $Master$ component first checks whether the $Actor$ components that have been registered to the framework have the corresponding resources to process the application. If true, the IoT application which contains one or multiple tasks will be scheduled; otherwise, the $Master$ component will inform the $User$ component that the current application cannot be processed. For each task of an IoT application, the scheduler will place it to the target $Actor$ component for execution based on Algorithm \ref{ppo}. After that, the $Actor$ component sends the relevant information (i.e., CPU utilization, RAM utilization, etc.) to the $Master$ component, which is stored in the buffer $\mathcal{D}_A$. The $User$ component also sends relevant information (i.e., response time, the result of task execution, etc.) to the $Master$ component, which is stored in the buffer $\mathcal{D}_U$. When the $Master$ collects sufficient information, it will update the scheduler, where the data in $\mathcal{D}_A$ and $\mathcal{D}_E$ are used to compute the reward for each step, as discussed in Algorithm \ref{ppo} and Eq. \ref{eq:33}, \ref{eq:34}, \ref{eq:35}.

\section{Performance Evaluation}
\label{performance_evaluation}
In this section, we first describe the experimental setup and sample applications used in the evaluation. Then, we investigate the hyperparameters of DRLIS. Finally, we discuss the performance of DRLIS by comparing it with its counterparts.

\subsection{Experiment Setup}
We first give a short introduction about the experimental environment and describe the IoT applications used in the experiment. Next, the baseline algorithms used to compare with DRLIS are presented.

\subsubsection{\textbf{Experiment Environment}}
\label{experiment_environment}
As discussed in Section \ref{implementation}, we implemented a scheduler based on DRLIS in the FogBus2 framework, and we use this scheduler for evaluation. We consider a heterogeneous experimental environment consisting of IoT devices, resource-limited fog servers, and resource-rich cloud servers. To simulate the heterogeneous multi-cloud computing environment, we used two instances of Nectar Cloud infrastructure (Intel Xeon 2 cores @2.0GHz, 9GB RAM, and Intel Xeon 16 cores @2.0GHz, 64GB RAM) and one instance of AWS Cloud (AMD EPYC 2 cores @2.2GHz, 4GM RAM). In the fog computing environment, to reflect the heterogeneity of the servers, we used a Raspberry Pi 3B (Broadcom BCM2837 4 cores @1.2GHz, 1GB RAM), a MacBook Pro (Apple M1 Pro 8 cores, 16GB RAM), and a Linux virtual machine (Intel Core i5 2 cores @3.1GHz, 4GB RAM). In addition, the IoT devices are configured with 2 cores @3.2GHz and 4GB RAM. Furthermore, we profiled the average bandwidth (i.e., data rate) and latency between servers as follows: the latency between the IoT device and the cloud server is around 15ms, and the bandwidth is around 6MB/s, while the latency between the IoT device and the fog server is around 3ms, and the bandwidth is around 25MB/s. Also, both $w_1$ and $w_2$ are set to 0.5 in Eq. \ref{eq:20}, meaning that the importance of load balancing and response time are equal. 

\subsubsection{\textbf{Sample IoT Applications}}
\label{app}
We used four IoT applications for evaluating the performance of the scheduler based on DRLIS. All applications implement both real-time and non-real-time features. Real-time means that the application can receive live streams and non-real-time means that the application can receive pre-recorded video files. Specifically, applications follow a sensor-actuator architecture, with each application operating as a single data stream. Sensors (e.g., cameras) capture environmental information and process it into data patterns (e.g., image frames) that will be forwarded to surrogate servers for processing, while actuators receive the processed data and represent the final outcome to the user. In addition, all applications provide a parameter called \textit{application label}, which can be used to set the frame size in the video. These applications are described as follows:
\begin{itemize}
\item \textit{Face Detection} \cite{goudarzi2021resource}: Detects and captures human faces. The human faces in the video are marked by squares. This application is implemented based on OpenCV\footnote{https://github.com/opencv/opencv\label{opencv}}.
\item \textit{Color Tracking} \cite{goudarzi2021resource}: Tracks colors from video. The user can dynamically configure the target colors through the GUI provided by the application. This application is implemented based on OpenCV\footref{opencv}.
\item \textit{Face And Eye Detection} \cite{goudarzi2021resource}: In addition to detecting and capturing human faces, the application also detects and captures human eyes. This application is implemented based on OpenCV\footref{opencv}.
\item \textit{Video OCR} \cite{deng2021fogbus2}: Recognizes and extracts text information from the video and transmits it back to the user. The application will automatically filter out keyframes. This application is implemented based on Google’s Tesseract-OCR Engine\footnote{https://github.com/tesseract-ocr/tesseract}.
\end{itemize}

\subsubsection{\textbf{Baseline Algorithms}}
To evaluate the performance of DRLIS, three other schedulers based on metaheuristic algorithms and reinforcement learning techniques are implemented, as follows:
\begin{itemize}
\item \textit{DQN}: It is one of the most adapted techniques in deep reinforcement learning, which constructs an end-to-end architecture from perception to decision. This algorithm has been used by many works in the current literature such as \cite{jie2021dqn}, \cite{9060882}, \cite{8657791}, and \cite{huang2019deep}. To compare with our proposed algorithm, we implement a DQN-based scheduler and integrate it into the FogBus2 framework. This scheduler can minimize the weighted load balancing and response time cost.
\item \textit{Q-Learning}: This technique belongs to value-based reinforcement learning techniques that combine the Monte Carlo method and the TD method. Its ultimate goal is to learn a table (\textit{Q-Table}). Works including \cite{baek2019managing}, \cite{aljanabi2021improving} adopt this technique. To integrate it into the FogBus2 framework, we implemented a scheduling policy. Furthermore, as a comparison, the scheduler can be used in the weighted cost problem to minimize the weighted load balancing and response time cost.
\item \textit{NSGA2}: It is a weighted cost genetic algorithm. It adopts the strategy of fast non-dominated sorting and crowding distance to reduce the complexity of the non-dominated sorting genetic algorithm. The algorithm has high efficiency and fast convergence rate \cite{yliniemi2016multi}. This algorithm is implemented using Pymoo \cite{9078759}.
\item \textit{NSGA3}: The framework of NSGA3 is basically the same as NSGA2, using fast non-dominated sorting to classify population individuals into different non-dominated fronts, and the difference mainly lies in the change of selection mechanism. Compared with NSGA2 using crowding distance to select individuals of the same non-dominated level, NSGA3 introduces well-distributed reference points to maintain population diversity under high-dimensional goals \cite{li2021achievement}. This algorithm is implemented using Pymoo \cite{9078759}.
\end{itemize}

\subsection{Hyperparameter Tuning}
\begin{figure*}[h]
\begin{subfigure}{.49\textwidth}
  \centering
  \includegraphics[width=\linewidth, height=5cm]{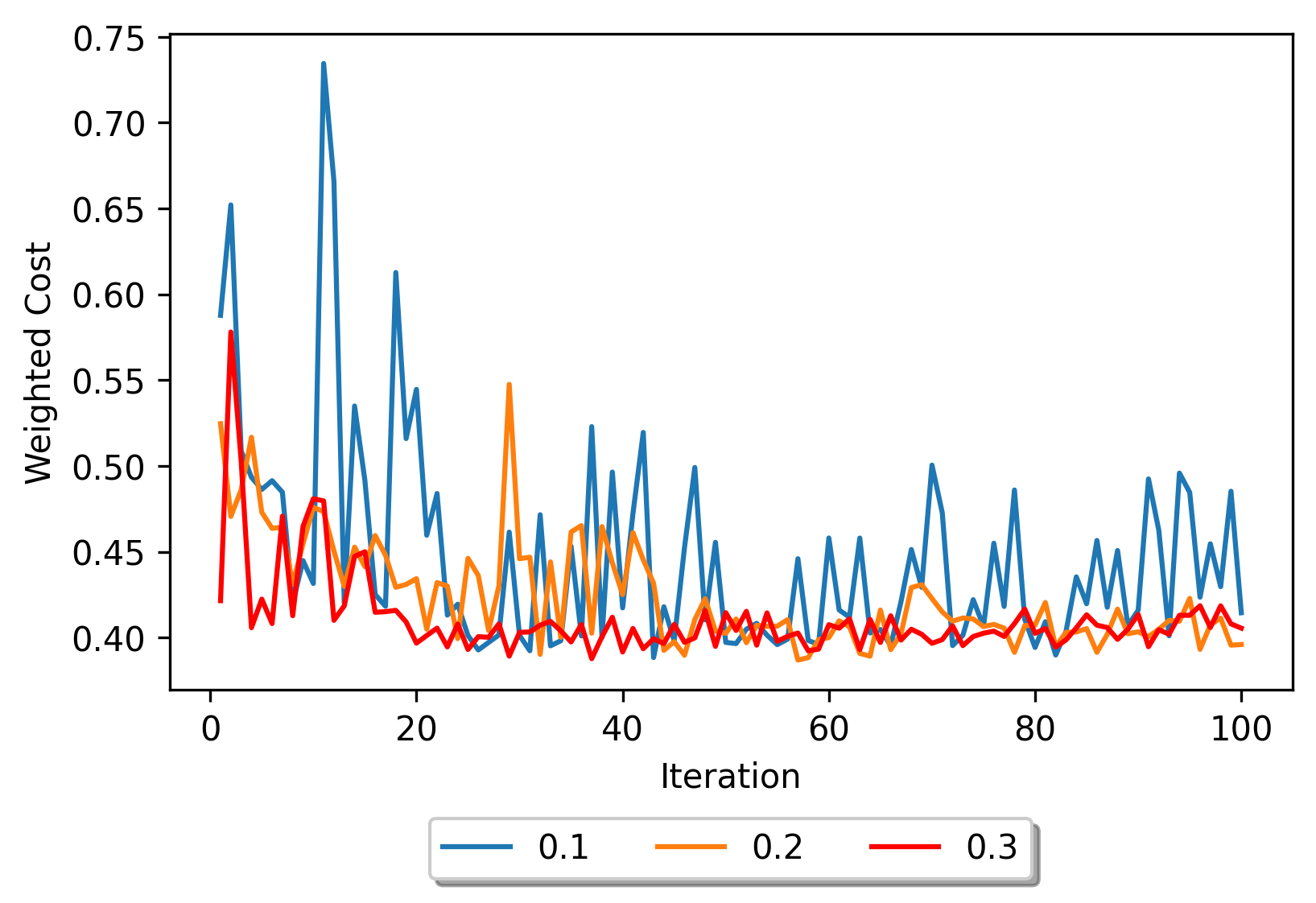}
  \caption{Clipping range}
  \label{fig:clippling}
\end{subfigure}%
\begin{subfigure}{.49\textwidth}
  \centering
  \includegraphics[width=\linewidth, height=5cm]{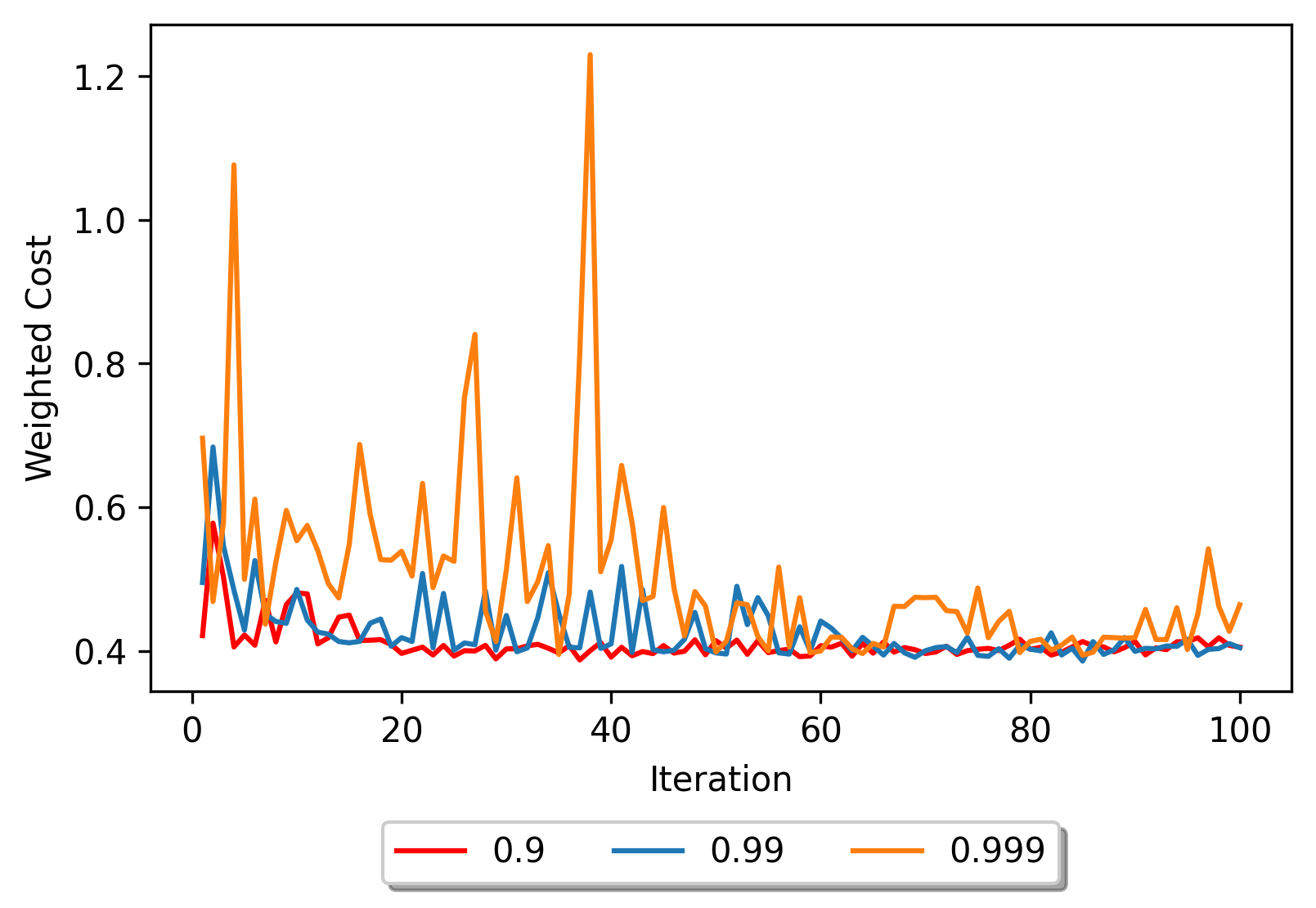}
  \caption{Discount factor}
  \label{fig:df}
\end{subfigure}
\begin{subfigure}{.49\textwidth}
  \centering
  \includegraphics[width=\linewidth, height=5cm]{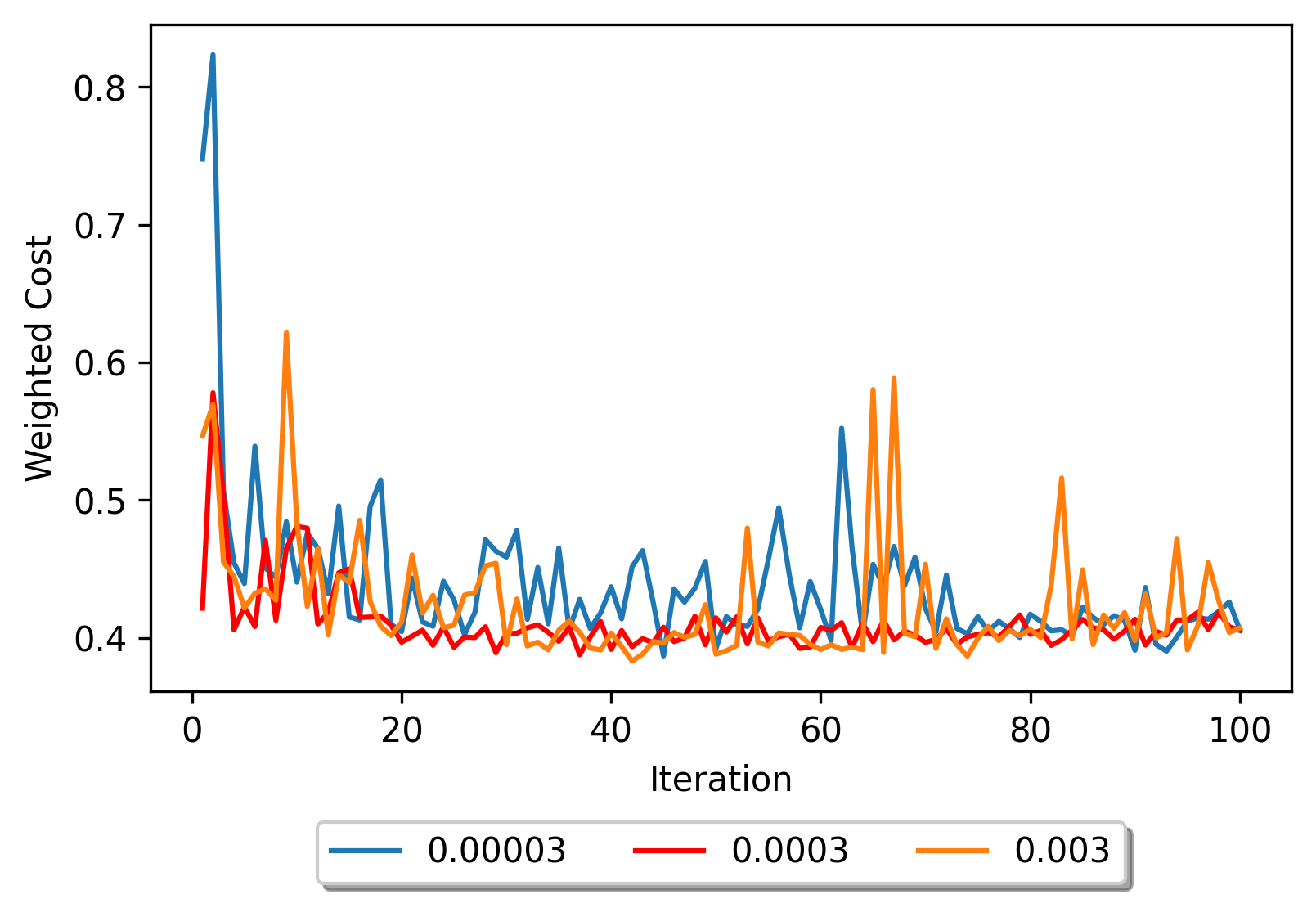}
  \caption{Actor network learning rate}
  \label{fig:alr}
\end{subfigure}
\begin{subfigure}{.49\textwidth}
  \centering
  \includegraphics[width=\linewidth, height=5cm]{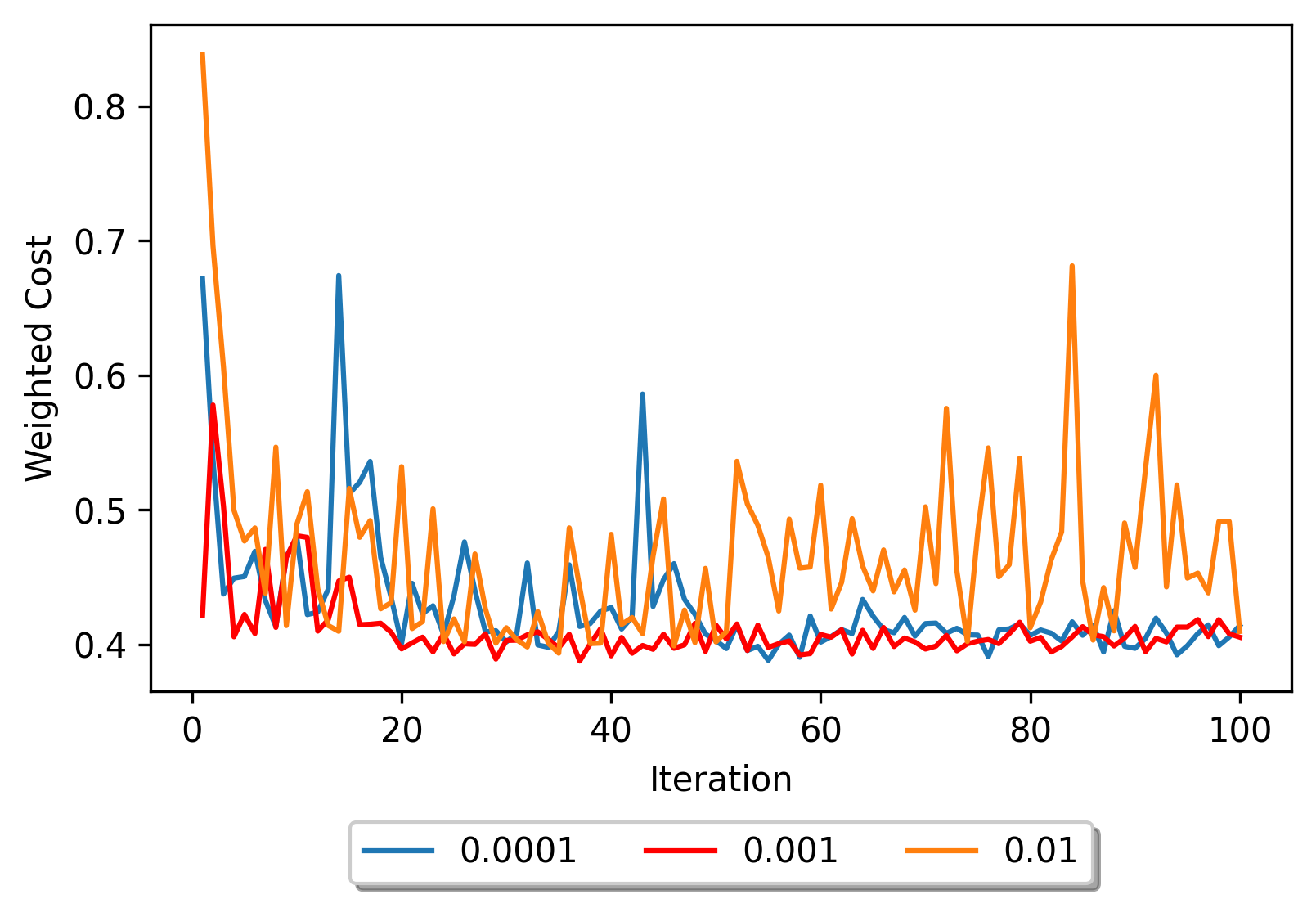}
  \caption{Critic network learning rate}
  \label{fig:clr}
\end{subfigure}
\caption{Hyperparameters tuning results}
\label{fig:pt}
\end{figure*}

The scheduler based on DRLIS is implemented via PyTorch. Considering the limited computational resources of some devices in the fog computing environment, both actor network and critic network consist of an input layer, a hidden layer, and an output layer. Henderson et al. \cite{henderson2018deep} investigate the effect of hyperparameter settings on the performance of reinforcement learning models. They survey the literature on different reinforcement learning techniques, list the hyperparameter settings used in the literature, and compare the actual performance of the models under different hyperparameter settings. They compare the performance of the PPO algorithm under different network architectures and the result shows that the model performs best under the network architecture where the hidden layer contains 64 hidden units and the hyperbolic tangent (TanH) function is used as the activation function. Therefore, we used the same network architecture for our experiments. In addition, we performed a grid search to tune the four main hyperparameters (i.e., clipping range, discount factor, learning rate for actor network, and learning rate for critic network), and the results are shown in Fig. \ref{fig:pt} The load balancing model control parameters $a1$ and $a2$ are both set to 0.5 to show the equal importance of CPU and RAM, however, these values can be tuned by users based on the objectives.

All the experiments regarding hyperparameters tuning are conducted in order to solve the weighted cost problem, as discussed in section \ref{mo_model}. We describe the process of hyperparameters tuning of our reinforcement learning model. For tuning the clipping range $\epsilon$, we followed Schulman et al. \cite{schulman2017proximal}, who proposed PPO and described that the model performs best with settings of clipping range $\epsilon$ among 0.1, 0.2, and 0.3. Fig. \ref{fig:clippling} shows that our model performs best when the clipping range $\epsilon$ is set to 0.3. For the discount factor $\gamma$, we reviewed related work on DRL in order to understand the common range for $\gamma$. According to \cite{schulman2017proximal, wei2022personalized}, the best setting for $\gamma$ sits somewhere among \{0.9-0.999\}. Accordingly, to keep the search area for tuning $\gamma$ in a viable range, we used the nominated values in these works and found that our model converges faster when $\gamma$ is set to 0.9. Fig.~\ref{fig:df} shows the tuning process of $\gamma$. Based on the similar approach for tuning $\epsilon$ and $\gamma$, for tuning the actor network learning rate $lr_a$, we referred to \cite{schulman2017proximal, henderson2018deep, bjorck2018understanding} for designing our tuning range. Accordingly, we used 0.003, 0.0003, and 0.00003 to tune $lr_a$. Fig. \ref{fig:alr} shows that our model performs best when the $lr_a$ is set to 0.0003. Considering the same approach for tuning, we followed \cite{huang2020reconfigurable, 9163056, islam2017reproducibility} and set our tuning range among \{0.01, 0.001, 0.0001\} and found that our model works best when $lr_c$ is 0.001. Fig. \ref{fig:clr} shows the performance of our model under different settings for $lr_c$. Overall, the deep neural network and training hyperparameters setting is presented in Table \ref{table:hyperparameters}. Besides, we also tune the hyperparameters for baseline techniques to fairly study their performance. The corresponding results are shown in Table \ref{table:hyperparameters_baseline}.

\begin{table}[]
\footnotesize
\caption{The hyperparameters setting for DRLIS}
\label{table:hyperparameters}
\centering
\begin{tabular}{|l|l|}
\hline
\textbf{DRLIS Hyperparameter}                       & \textbf{Value} \\ \hline
Neural Network Layers                            & 3              \\ \hline
Hidden Layer Units                       & 64             \\ \hline
Optimization Method                      & Adam           \\ \hline
Activation Function                      & TanH           \\ \hline
Clipping Range $\epsilon$                           & 0.3            \\ \hline
Discount Factor $\gamma$                          & 0.9            \\ \hline
Actor Learning Rate $lr_a$                     & 0.0003         \\ \hline
Critic Learning Rate $lr_c$                    & 0.001          \\ \hline
Policy Objective Function Coefficient $a_c$    & 1              \\ \hline
Value Function Loss Function Coefficient $a_v$ & 0.5            \\ \hline
Entropy Bonus Coefficient $a_e$                & 0.01           \\ \hline
Load Balancing Model CPU Control Parameter $a_1$ &0.5           \\ \hline
Load Balancing Model RAM Control Parameter $a_2$ &0.5           \\ \hline
\end{tabular}%

\end{table}

\begin{table}[]
\footnotesize
\caption{The hyperparameters setting for baseline techniques}
\label{table:hyperparameters_baseline}
\centering
\begin{tabular}{|l|l|}
\hline
\textbf{DQN Hyperparameter}                       & \textbf{Value} \\ \hline
Neural Network Layers                            & 3              \\ \hline
Hidden Layer Units                       & 64             \\ \hline
Optimization Method                      & Adam           \\ \hline
Activation Function                      & ReLU           \\ \hline
Discount Factor                         & 0.99            \\ \hline
Learning Rate                     & 0.0001         \\ \hline
Exploration Rate                      & 1           \\ \hline
Exploration Decay                      & 0.9          \\ \hline
Minimum Exploration                       & 0.05           \\ \hhline{|==|}

\textbf{Q-Learning Hyperparameter}                       & \textbf{Value} \\ \hline
Discount Factor                       & 0.9             \\ \hline
Learning Rate                     & 0.1         \\ \hhline{|==|}
\textbf{NSGA2 and NSGA3 Hyperparameter}                       & \textbf{Value} \\ \hline
Population Size                      & 200             \\ \hline
Generation Numbers                    & 100         \\ \hline
\end{tabular}%
\end{table}

\subsection{Performance Study}
We performed two experiments to evaluate DRLIS compared to its counterparts, regarding the load balancing of the servers, the response time of the IoT applications, and the weighted cost.
\subsubsection{Cost vs Policy Update Analysis}
\label{cost}
\begin{figure*}[h]
\begin{subfigure}{0.33\textwidth}
  \centering
  \includegraphics[width=\linewidth]{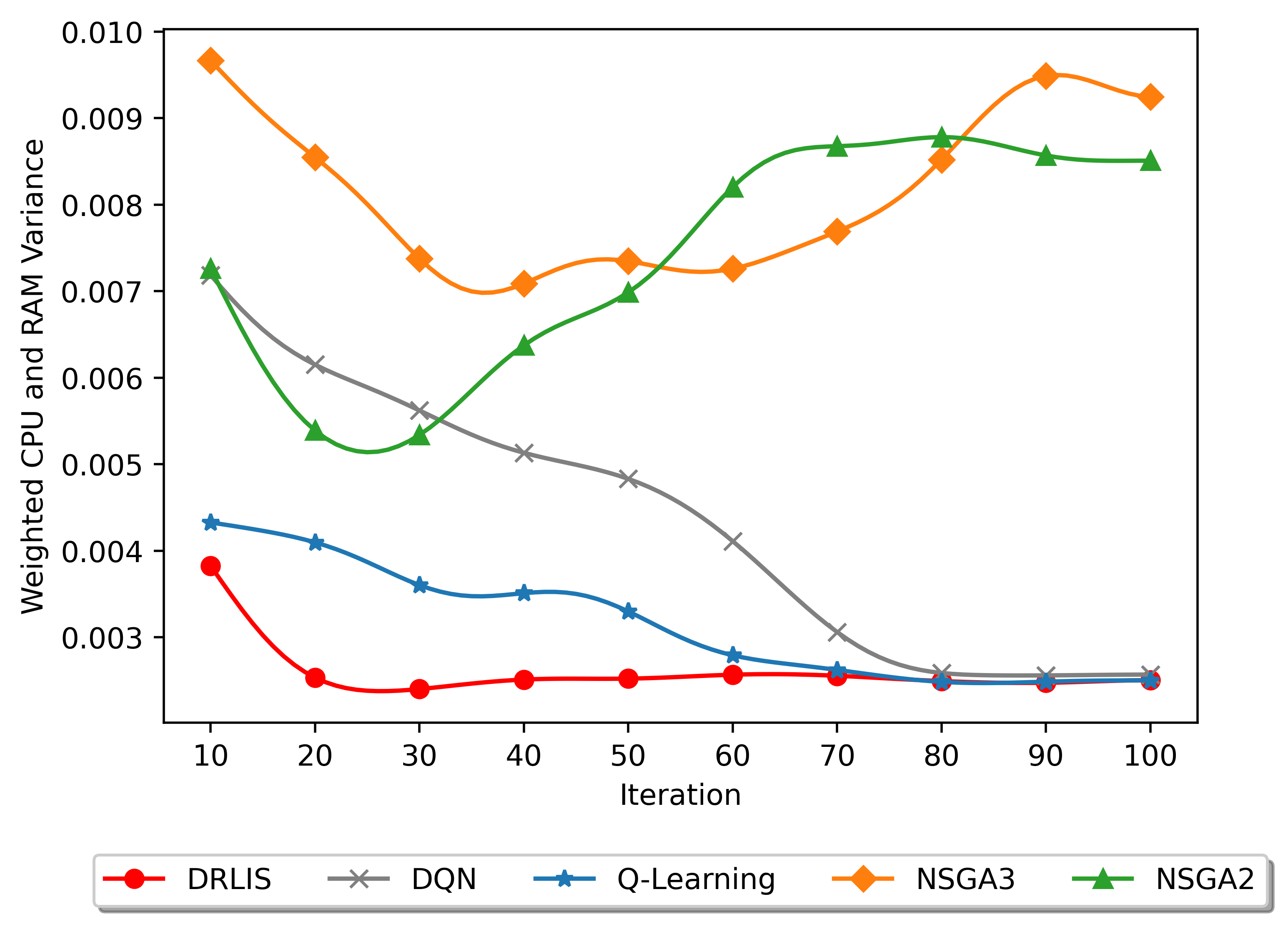}
  \caption{Load balancing}
  \label{fig:lbt}
\end{subfigure}%
\begin{subfigure}{0.33\textwidth}
  \centering
  \includegraphics[width=\linewidth]{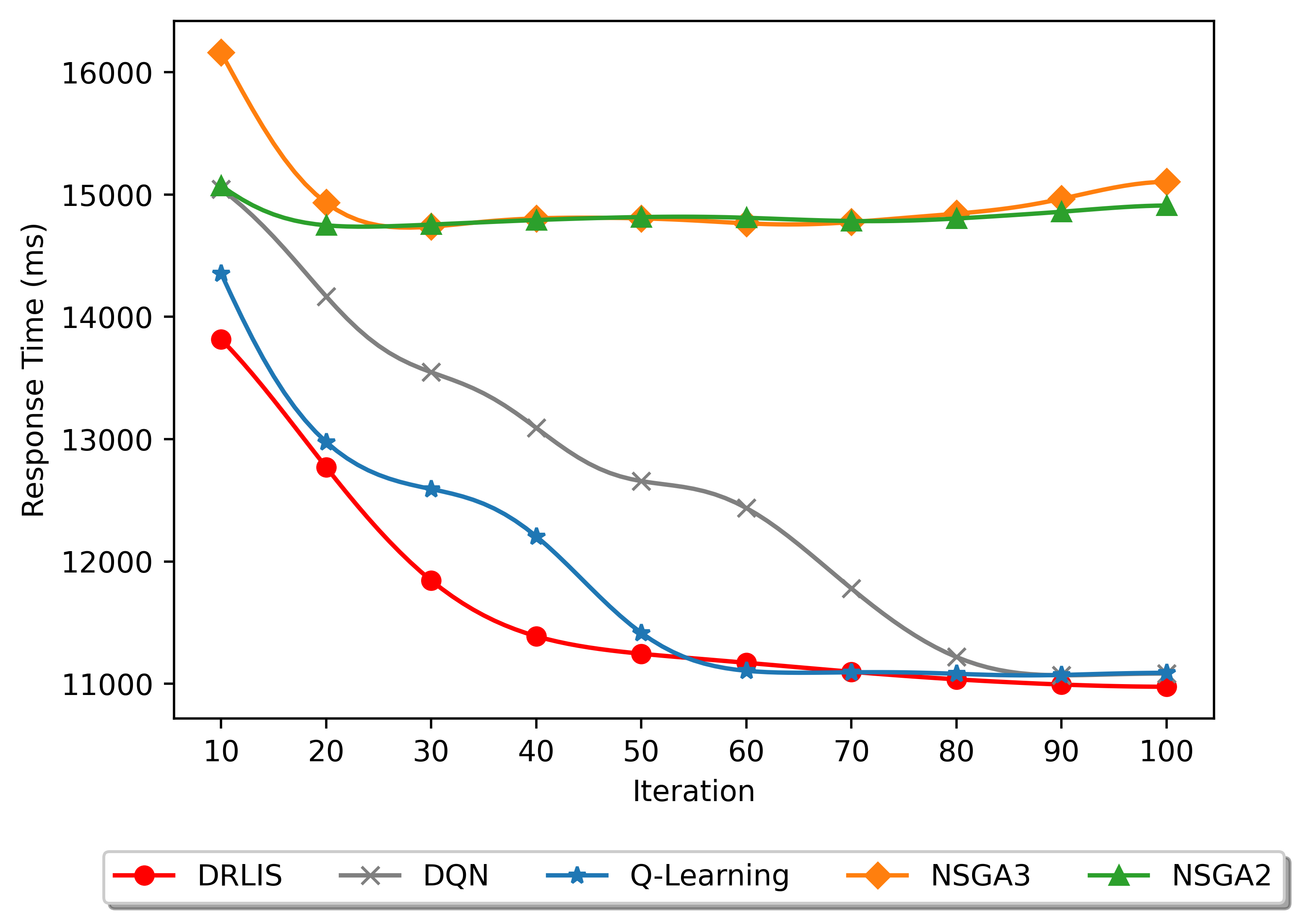}
  \caption{Response time}
  \label{fig:rtt}
\end{subfigure}
\begin{subfigure}{0.33\textwidth}
  \centering
  \includegraphics[width=\linewidth]{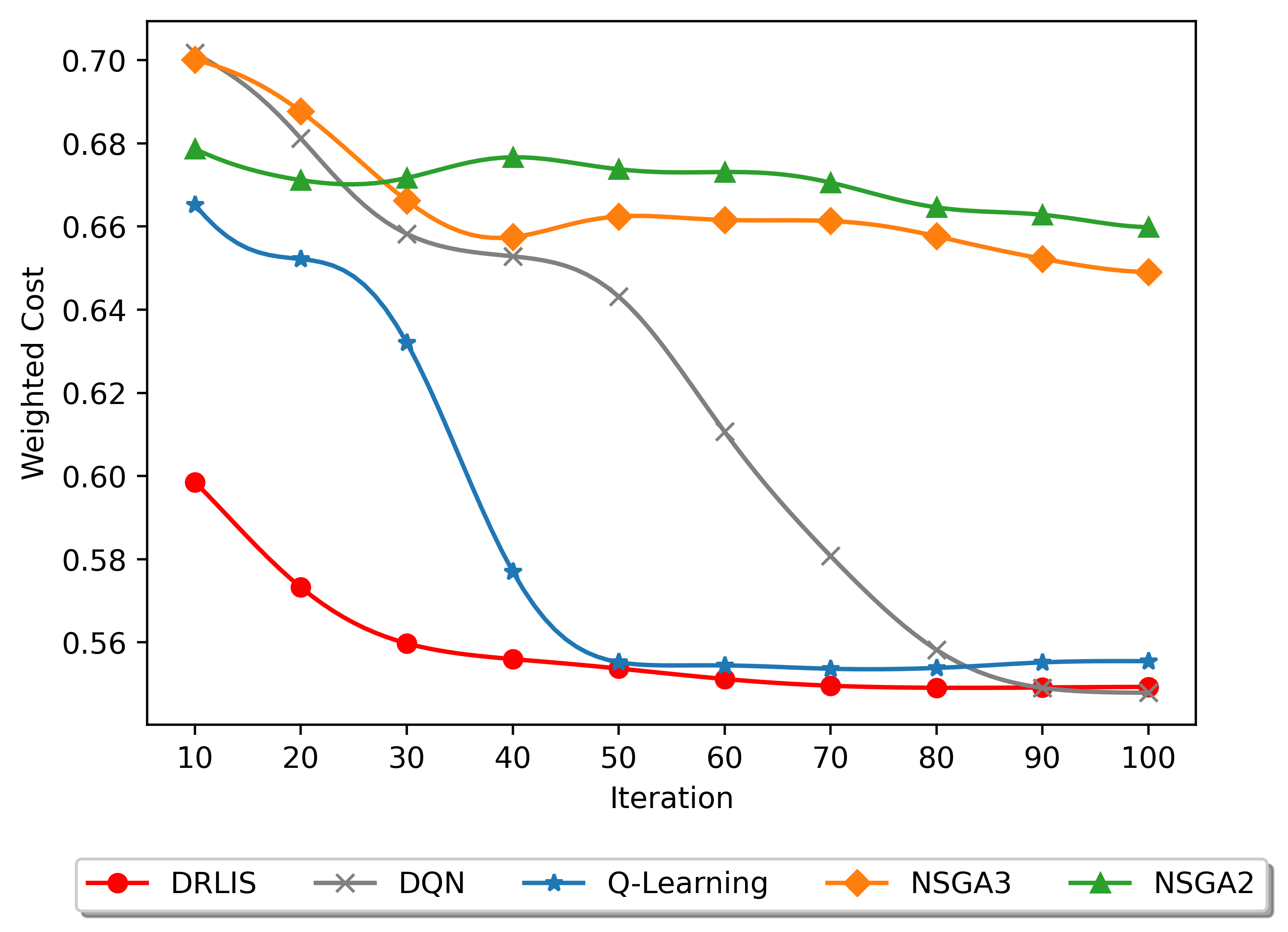}
  \caption{Weighted cost}
  \label{fig:mot}
\end{subfigure}
\caption{Cost vs policy update analysis - train phase}
\label{fig:train}
\end{figure*}

In this experiment, we investigate the algorithm performance in different iterations when the policy is updated. We used the four applications mentioned in Section \ref{app} for training with the resolution parameter set to 480, and the maximum number of iterations is set to 100. The training results of algorithms with the three optimization objectives are shown in Fig. \ref{fig:train}.

As shown in Fig. \ref{fig:lbt}, when optimizing the load-balancing problem of the servers, the average computational resource variance of the servers is lower for the Q-Learning-based, DQN-based, and DRLIS-based schedulers than for the NSGA2-based and NSGA3-based schedulers. Moreover, only the reinforcement learning-based scheduler can achieve a stable convergence state. However, the Q-Learning-based scheduler requires more than 60 iterations before reaching a converged state, and the DQN-based scheduler requires more than 80 iterations, while the DRLIS-based scheduler only requires about 20 updates to converge to a similar stable state. The NSGA2-based and NSGA3-based schedulers are unable to reach the convergence state. As shown in Fig. \ref{fig:rtt}, when optimizing the response time problem of the application, unlike the former problem, all schedulers can converge. However, the average response time of Q-Learning-based, DQN-based, and DRLIS-based schedulers is still lower than that of NSGA2-based and NSGA3-based schedulers. In addition, the DRLIS-based scheduler still outperforms the Q-Learning-based and the DQN-based schedulers in terms of convergence speed. Finally, as Fig. \ref{fig:mot} shows, when optimizing the weighted cost problem, similar to the load balancing problem, the average cost is lower for the Q-Learning-based, DQN-based, and DRLIS-based schedulers than for the NSGA2 and NSGA3-based schedulers, and only the first three can reach a stable convergence state. Moreover, although the Q-Learning-based, DQN-based, and DRLIS-based schedulers have similar final convergence levels, the DRLIS-based scheduler converges much faster than the Q-Learning-based and DQN-based schedulers. This proves that the DRLIS-based scheduler outperforms the other techniques in terms of average cost, convergence, and convergence speed during the training phase.

\begin{figure*}[t]
\begin{subfigure}{0.33\textwidth}
  \centering
  \includegraphics[width=\linewidth]{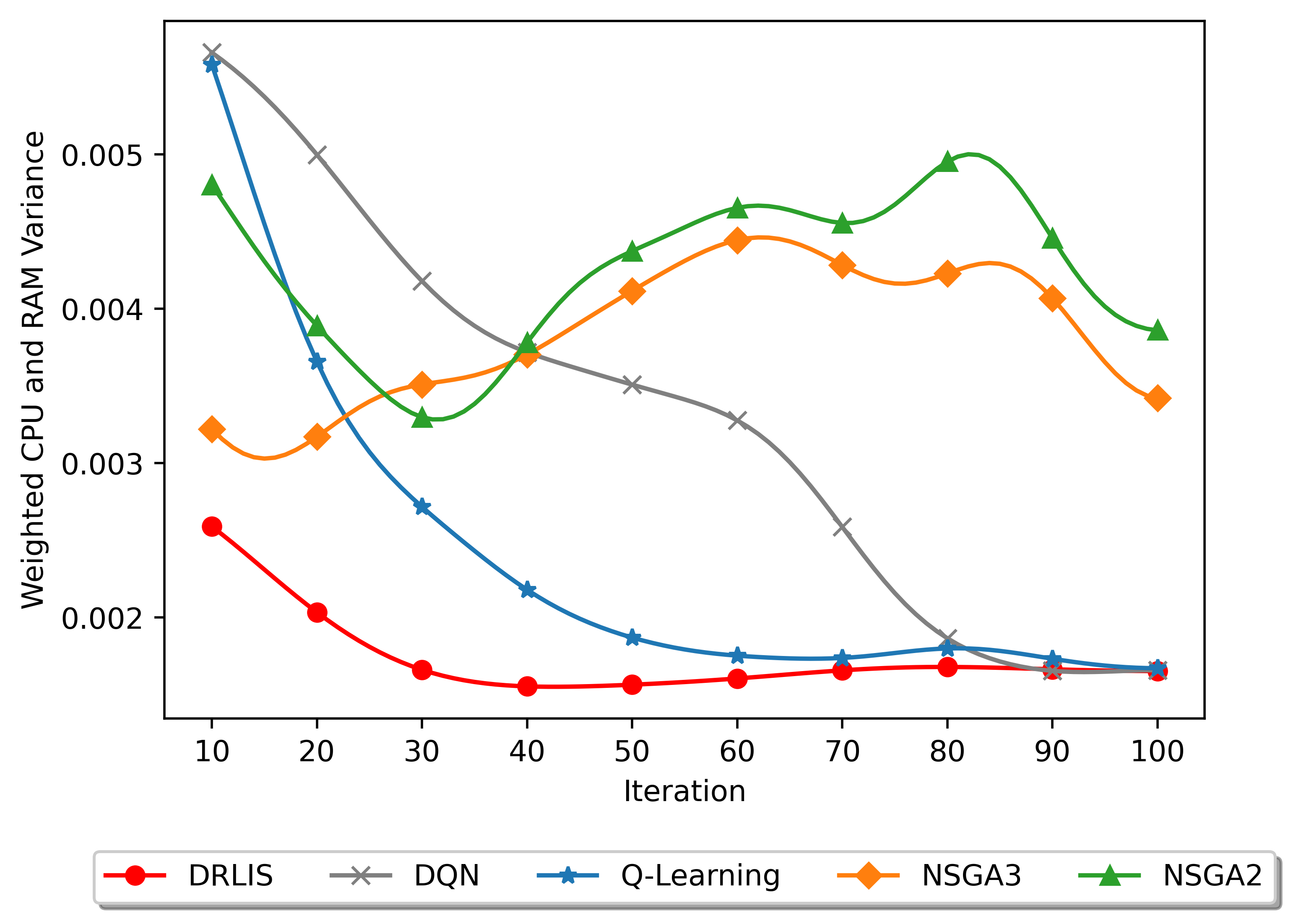}
  \caption{Load balancing}
  \label{fig:lbe}
\end{subfigure}%
\begin{subfigure}{0.33\textwidth}
  \centering
  \includegraphics[width=\linewidth]{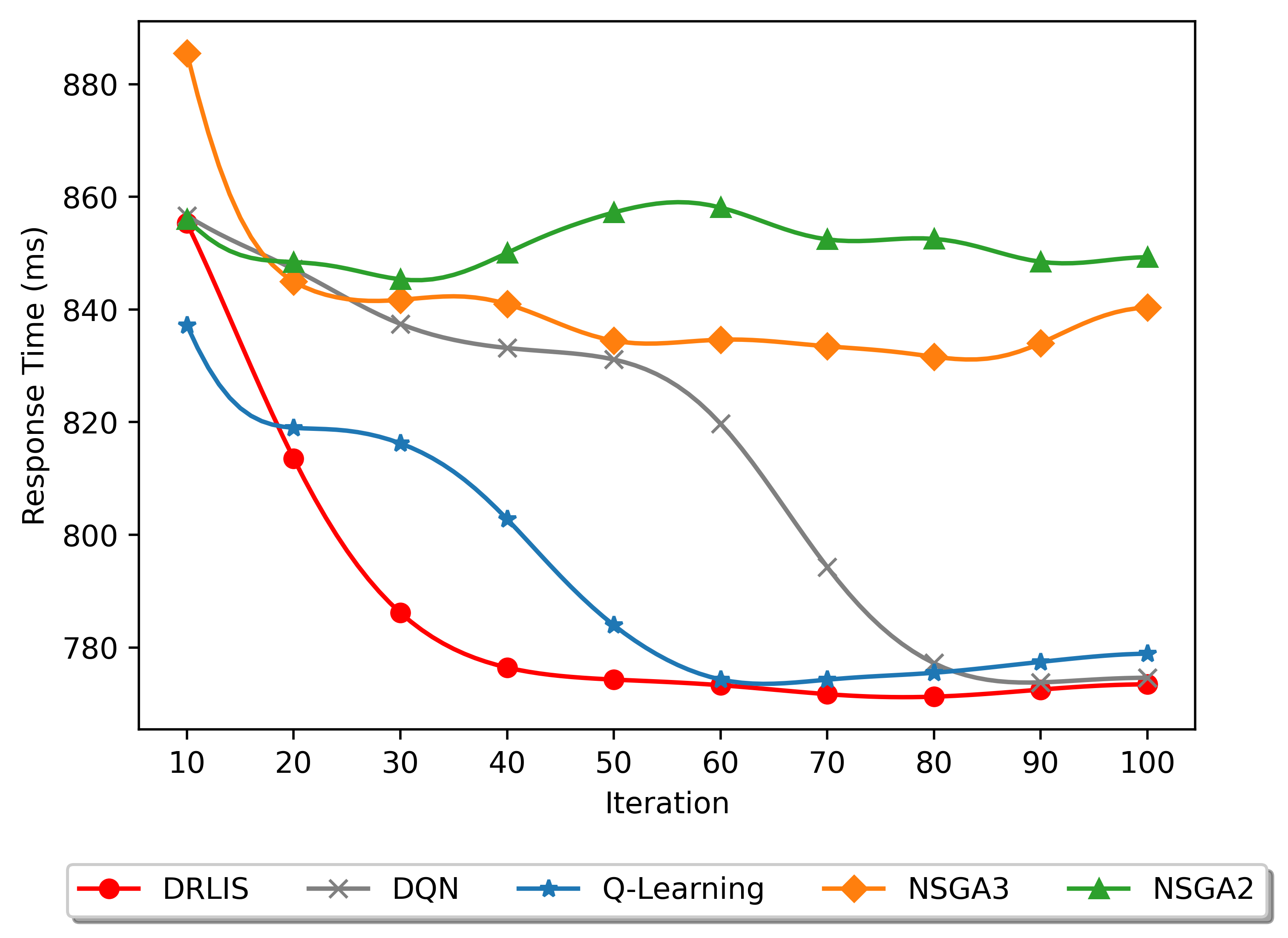}
  \caption{Response time}
  \label{fig:rte}
\end{subfigure}
\begin{subfigure}{0.33\textwidth}
  \centering
  \includegraphics[width=\linewidth]{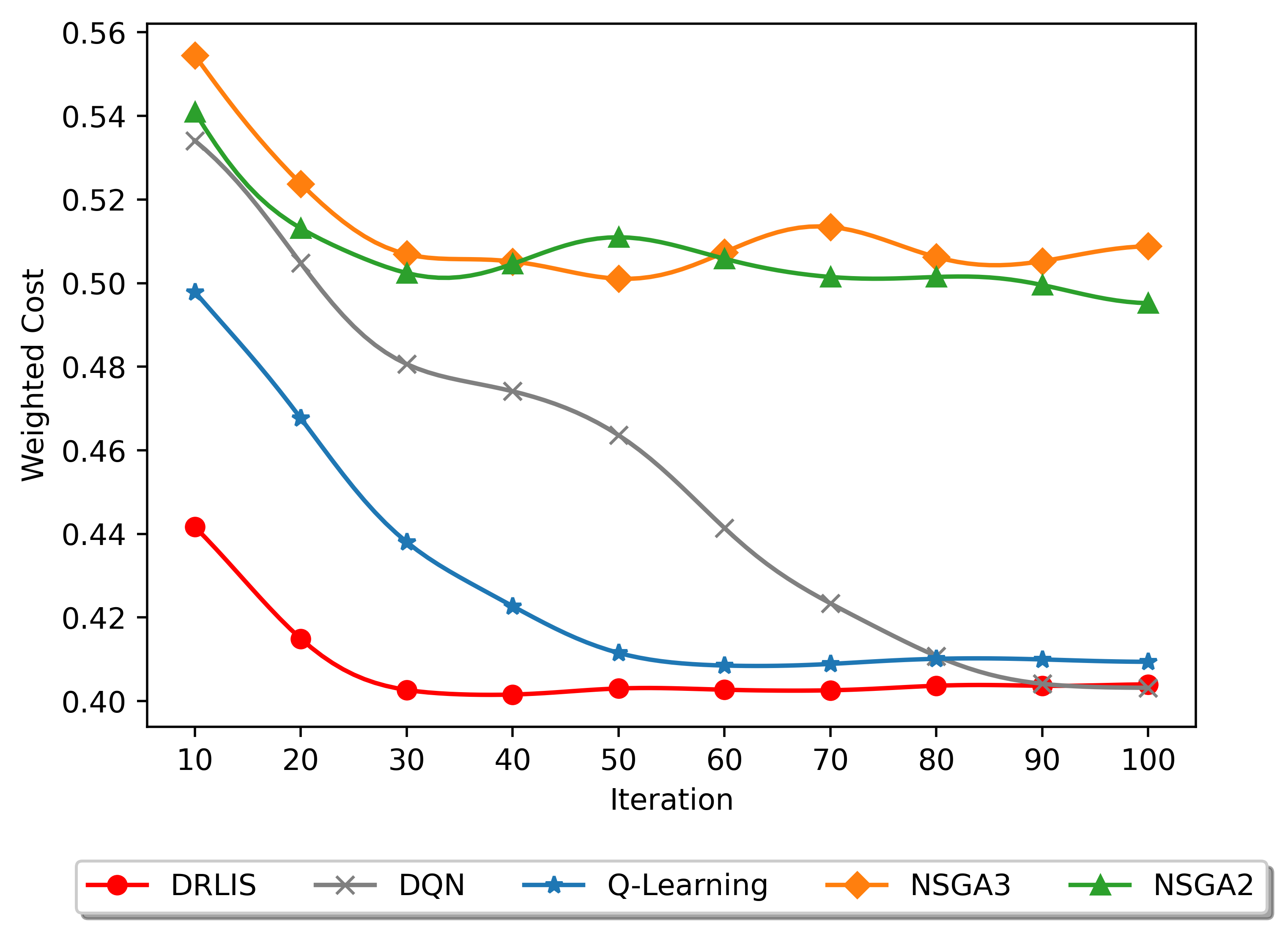}
  \caption{Weighted cost}
  \label{fig:moe}
\end{subfigure}
\caption{Cost vs policy update analysis - evaluation phase}
\label{fig:eva}
\end{figure*}

In the evaluation phase, we set the resolution to 240, which will make the demand for computational resources and response time of the IoT application different from the training phase. The evaluation phase results of the different algorithms regarding the three optimization objectives are shown in Fig. \ref{fig:eva}. It can be observed that when the optimization objective is server load balancing, IoT application response time, and weighted cost, respectively, the schedulers based on different algorithms have similar performances as the training phase. Specifically, only the cost of the Q-Learning-based, DQN-based, and DRLIS-based schedulers converges, and the cost of the NSGA2-based and NSGA3-based schedulers fluctuates up and down in a higher range. Moreover, the average and final costs of the Q-Learning-based, DQN-based, and DRLIS-based schedulers are significantly lower than those of the NSGA2-based and NSGA3-based schedulers during the evaluation phase. In addition, in the weighted cost scenario, the DRLIS-based scheduler can converge the cost to a stable level after about 30 policy updates, while the Q-Learning-based scheduler usually takes about 60 updates to converge to a slightly higher level, and the DQN-based scheduler needs more than 80 updates to converge to the same level. Overall, compared with the Q-Learning-based scheduler, which can converge stably and with the fastest convergence speed in the baseline algorithms, the average performance of the DRLIS-based scheduler improves by 55\%, 37\%, and 50\%, in terms of servers load balancing, IoT application response time, and weighted cost, respectively.

\subsubsection{Scheduling Overhead Analysis}
In this section, we investigate the scheduling overhead of different techniques-based schedulers when handling IoT applications. The environment settings are the same as Section \ref{experiment_environment}, and the resolution of the IoT applications is set to 480. For each scheduler, we repeat the experiment for 100 rounds, feeding four IoT applications to the scheduler in each round. Besides, we define the average scheduling overhead as $T_{ave} = \frac{T_{total}}{100}$, where $T_{total}$ represents the total overhead spent by the scheduler to handle the applications in 100 rounds. 

\begin{figure}[!htb]
\centering
\includegraphics[width=\linewidth]{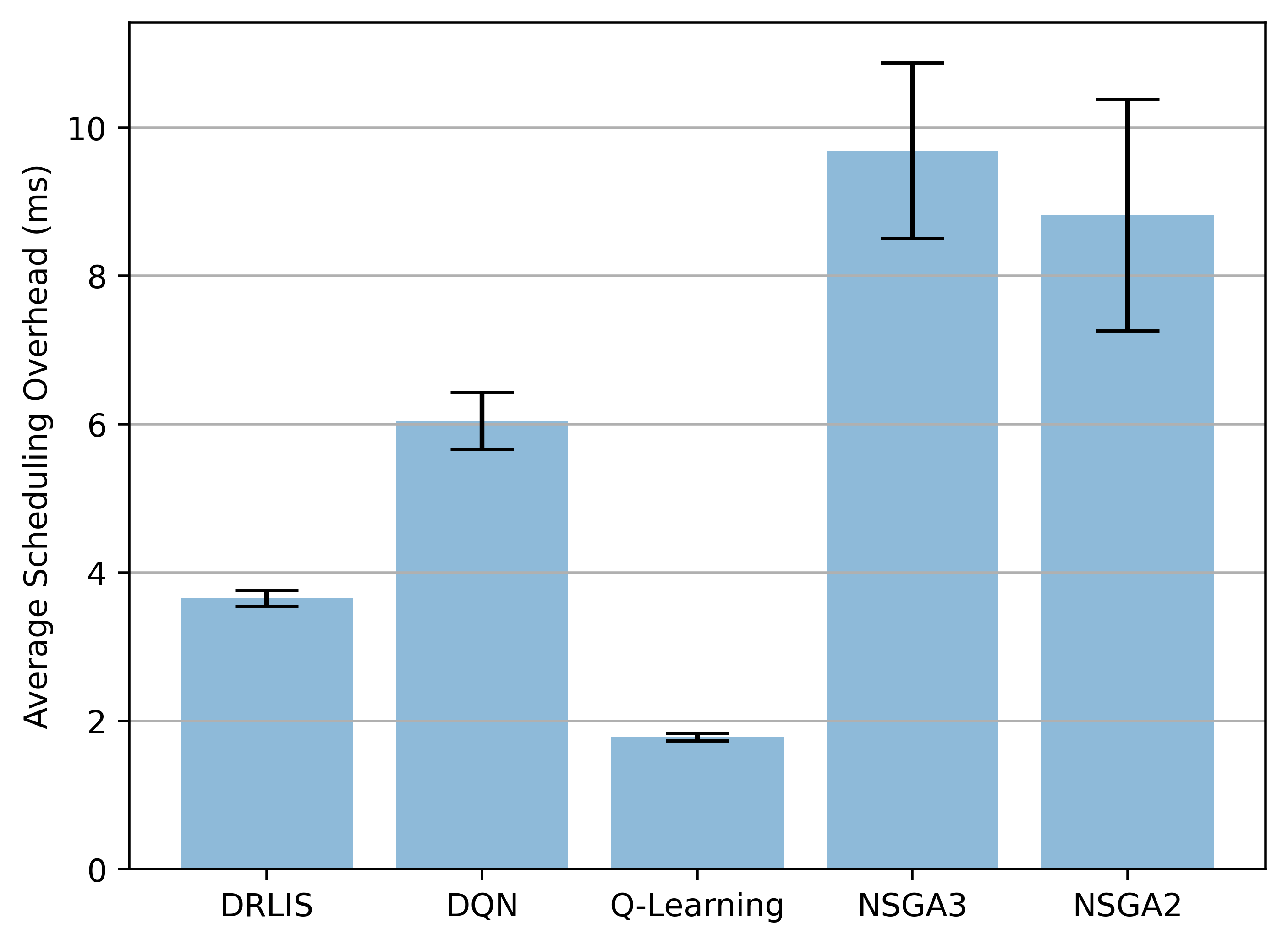}
\caption{Average scheduling overhead with a 95\% CNFI}
\label{fig:aso}
\end{figure}

Figure \ref{fig:aso} depicts the average scheduling overhead $T_{ave}$ with a 95\% Confidence Interval (CNFI) of schedulers based on different technologies when handling IoT applications. It is obvious that the scheduling overheads of reinforcement learning techniques (i.e., DRLIS, DQN, Q-Learning) are usually lower than metaheuristics techniques (i.e., NSGA2, NSGA3). In addition, the 95\% CNFI of the scheduling overhead of reinforcement learning techniques is also much shorter than metaheuristic techniques. Specifically, the scheduling overhead of DRLIS is more than 50\% lower than NSGA2 and NSGA3, and more than 33\% lower than DQN, but it is about 2ms more than Q-Learning. However, considering that the convergence speed of DRLIS is much faster than that of Q-Learning, as discussed in Section \ref{cost}, the increased overhead cost of DRLIS over Q-Learning can be negligible. Therefore, in the heterogeneous edge and fog computing environment, our proposed DRLIS-based algorithm can handle the weighted cost optimization problem of IoT applications more efficiently than other techniques.

\section{Conclusions and Future Work}
\label{conclusion}
In this paper, we proposed DRLIS, a DRL-based algorithm to solve the weighted cost optimization problem for IoT applications scheduling in heterogeneous edge and fog computing environments. First, we proposed corresponding cost models for optimizing load balancing and response time in heterogeneous edge and fog computing environments and formulate a weighted cost model based on both of them. In addition, we implemented a practical scheduler in the FogBus2 function-as-a-service framework for scheduling IoT applications. Compared to existing work, DRLIS has significant advantages in convergence speed, optimization cost, and scheduling overhead. Through extensive experiments and comparisons with other works in the literature, DRLIS achieves performance improvements of up to 49\%, 60\%, and 55\% in terms of load balancing, response time, and weighted cost, respectively.

For future work, considering the limited resources and the distribution of the devices in edge computing, we plan to explore distributed deep reinforcement learning to further improve the scheduler's performance. Also, we plan to consider more models to extend our proposed weighted cost model, including economic aspects and energy consumption aspects in large-scale serverless computing environments. In addition, to optimize the performance of IoT applications involving GPU tasks (e.g., image processing oriented applications), we will extend FogBus2 framework to consider resource usage when scheduling such applications on Application-Specific Integrated Circuit (ASIC)/GPU-based edge and cloud servers for more efficient performance.

\bibliographystyle{elsarticle-num}


\end{document}